\documentclass{jfm}
\usepackage{graphicx}
\usepackage{background}
\backgroundsetup{contents=, opacity=0.25}       

\usepackage{xcolor} 

\shorttitle{Pressure-driven Stokes flow along finite slip grooves in tubes and annuli}
\shortauthor{S. Zimmermann and C. Schönecker}

\title{Analytical models for pressure-driven Stokes flow through superhydrophobic and liquid infused tubes and annular pipes}

\author{Sebastian Zimmermann\aff{1}
\and Clarissa Schönecker\aff{1}
 \corresp{\email{c.schoenecker@mv.rptu.de}}}

\affiliation{\aff{1}Rheinland-Pfälzische Technische Universität Kaiserslautern-Landau (RPTU), D-67663 Kaiserslautern, Germany}

\begin{document}
\maketitle

\begin{abstract}
Analytical expressions for the velocity field and the effective slip length of pressure-driven Stokes flow through slippery pipes and annuli with rotationally symmetrical longitudinal slits are derived. Specifically, the developed models incorporate a finite local slip length or shear stress along the slits and thus go beyond the assumption of perfect slip commonly employed for superhydrophobic surfaces. Thereby, they provide the possibility to assess the influence of both the viscosity of the air or other fluid that is modelled to fill the slits as well as the influence of the micro-geometry of these slits. Firstly, expressions for tubes and annular pipes with superhydrophobic or slippery walls are provided. Secondly, these solutions are combined to a tube-within-a circular pipe scenario, where one fluid domain provides a slip to the other. This scenario is interesting as an application to achieve stable fluid-fluid interfaces. With respect to modelling, it illustrates the specification of the local slip length depending on a linked flow field. The comparisons of the analytically calculated solutions with numerical simulations shows excellent agreement. The results of this article thus represent an important instrument for the design and optimization of slippage along surfaces in circular geometries. 
\end{abstract}

\begin{keywords}

\end{keywords}

\section{Introduction}\label{sec:intro}
Flows over micro- or nanostructured surfaces are of great technical importance. For example, grooves, posts or holes are incorporated into no-slip walls whose structures contain a secondary immiscible fluid, as in the case of superhydrophobic surfaces \citep{shirtcliffe2010introduction}, where air is trapped between the primary water flow and the wall structuring. This wetting scenario is called the Cassie state, where the primary fluid wets only the upper surface of the structuring, forming a heterogeneous solid-liquid and solid-gas interface \citep{cassie1944_wettability}. Such configurations lead to a lower surface wettability and thus to a self-cleaning and water-repellent behavior as for the Lotus leaf \citep{koch2009multifunctional}. Furthermore, the relative surface fraction of the no-slip wall is reduced and the primary fluid can slide over air cushions on the surface, reducing drag significantly \citep{Rothstein_2010_uperhydrophobicsurfaces, Karatay_2013_Controlslippagetunable, ou2004laminar}. This is of great importance since hydrodynamic drag is directly related to the energy required to transport a fluid through a domain bounded by walls, implying a potential high economic incentive. Air is a suitable secondary fluid due to its low viscosity. However, air dissipation from the microstructures can occur. The interface then migrates into the microstructure (sagging) and drives the remaining air out. The result is the Wenzel state \citep{wenzel1936resistance}, i.e. the primary fluid has conquered and filled all microstructures. The slipping effect is then significantly reduced. Surfaces impregnated with a lubricant instead of air (liquid-infused surfaces - LIS) have proven to be a useful tool to prevent such Cassie-Wenzel transitions \citep{wong2011bioinspired} while still allowing slippage \citep{Asmolov_2018_Enhancedslipproperties} and are therefore receiving increasing attention \citep{Hardt_2022_Flowdroptransport}. LIS promote anti-icing \citep{latthe2019recent} and anti-biofouling \citep{epstein2012liquid} effects, which are major challenges in various industries such as transportation, agriculture and energy \citep{Ras_2016_Nonwettablesurfaces, cao2009anti-acing, agbe2020tunable}. To maximize the area fraction of the secondary fluid and to account for possible interface collapse, numerous microstructures are incorporated onto the surface.

\begin{figure}
    \centerline{\includegraphics{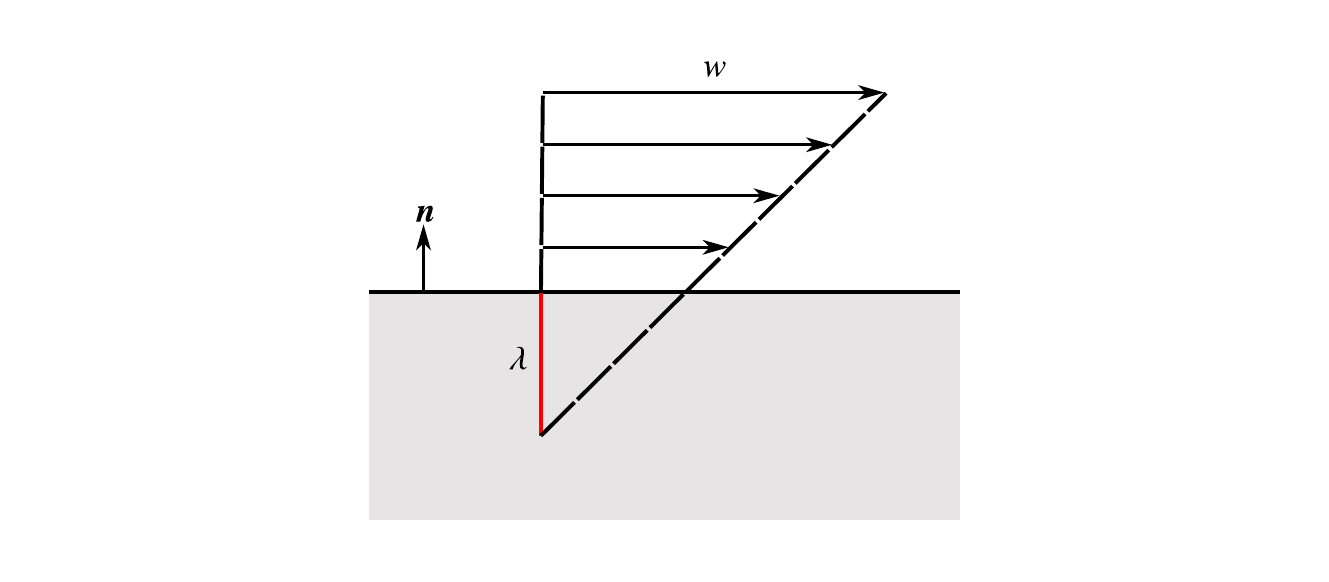}}
    \caption{Schematic illustration of the slip boundary condition}
    \label{fig:fig_slip_length}
\end{figure}

Since these are much smaller than the general geometry, complete numerical resolution of the full domain is not feasible. One way to solve this problem is to average the cumulative effect of all microstructures across the patterned wall and implement it into the numerical model using the Navier slip boundary condition
\begin{equation}
    w = \lambda_{\mathrm{eff}} \frac{\partial w}{\partial \boldsymbol{n}}.
    \label{eq: navier-slip-cond.}
\end{equation} 
The effective slip length $\lambda_{\mathrm{eff}}$ can be understood as an virtual depth below the surface where the velocity is extrapolated to zero (figure \ref{fig:fig_slip_length}) and connects the velocity $w$ with its normal derivative on the slip-wall. Thus, it is an important input quantity for numerical studies of microstructured surfaces and can be extracted from analytical models. 
Superhydrophobic structures of particular importance for application are parallel longitudinal stripes on pipe surfaces, see figure \ref{fig:connection_pipe_annular_inf-plates_geometry}. An important result on this subject is given by \citet{Philip_1972a} who derived an analytical solution for the pressure-driven flow field of a pipe containing $N$ rotationally symmetric no-shear wall sections on an otherwise no-slip wall. \citet{LaugaStone_2003} took up that formula and derived the effective slip length for such a configuration using \citep{Philip_1972b} 
\begin{equation}
    \tilde{\lambda}_{\mathrm{eff}} = \frac{\lambda_{\mathrm{eff}}}{R_0} = \frac{2}{N} \ln\left(\sec\left(\frac{\theta}{2} \right)  \right),
    \label{eq: effective-slip-length_philip}
\end{equation}  
with the pipe radius $R_0$, number of no-shear slits $N$ and the angle $\theta$ as a measure for the no-shear fraction with $\theta/ \pi$ being the proportion of the total wall surface occupied by no-shear slits. \citet{Crowdy_2021_theoStud} derives, among other things, analytical solutions for the effective slip length of pressure-driven annular flows containing longitudinal stripes on the inner or outer wall. The former is given by
\begin{equation}
    \tilde{\lambda}_{\mathrm{eff}} =  \tilde{R}_1 \ln(\tilde{R}_1) - 2 \tilde{R}_1 \frac{S}{I_{-1}},
    \label{eq: effective-slip-length_crowdy}
\end{equation}
with $\tilde{\lambda}_{\mathrm{eff}}, \tilde{R}_1$ in equation \ref{eq: effective-slip-length_crowdy} being normalized with respect to the outer radius of the annulus. $\tilde{R}_1$ denotes the dimensionless inner radius of the annulus. The outer wall is set to be the unit disk. $S$ is a scaling constant and $I_{-1}$ the coefficient in a Laurent expansion performed by \citet{Crowdy_2021_theoStud}, both solely dependent on geometric parameters. For further details, see section \S\ref{subsubsec:math_description_annulus_eff-slip}. However, both approaches assume perfect slip along certain boundary parts, i.e. the primary flow experiences no shear stress there. This is, of course, a condition which cannot be achieved in reality, but represents an ideal limit where the enclosed fluid is decoupled from the bulk flow. The local slip length is infinite. Thus, in such models, it does not matter whether the microstructures are impregnated by air, oil or water, the effect remains unchanged. In addition to the viscous interaction of both fluids, the influence of the microstructure geometry is also not considered, although it plays a crucial role \citep[see][]{Schoenecker_2013_singleCav, Schoenecker_2015_arrayCav}.  

To close this gap, this work derives analytic equations for the pressure-driven pipe flow field that have rotationally symmetric finite slip slits on a wall using a superposition approach. Classical pipe (section \S\ref{subsec:math_description_pipe}) and annular pipe flows (section \S\ref{subsec:math_description_annulus}) are considered, the latter having grooves on the inner wall. The quantity of the effective slip length as a function of a finite local slip length is given for both cases. Such equations are elementary for practical applications. These finite slip solutions are then linked so that a coupled velocity field of a pipe-within-pipe is obtained (section \S\ref{sec:coupling}). The local slip lengths act as a linking coefficient and depend on the connected flow of the other domain. This connection opens the computability of a wide range of possible applications. Finally, the results of this work are discussed, analyzed and illustrated in section \S\ref{sec:results-discussion}.

\section{Mathematical description of the flow field and effective slip length}\label{sec:math_description}
A pressure driven Stokes flow $(0,0,w(x,y))$ of a fluid of viscosity $\mu$ along the $Z$ axis of an arbitrary Cartesian domain $(x,y,Z)$ having a cross section in the $(x,y)$ plane is goverened by the Poisson equation 
\begin{equation}
    \nabla^2 w(x,y) = - \frac{s}{\mu},
    \label{governing_equation_poisson}
\end{equation}
where $-s$ is the negative pressure gradient along the $Z$ axis and $\nabla$ being the Nabla operator
\begin{equation}
    \nabla^2 = \frac{\partial^2}{\partial x^2} + \frac{\partial^2}{\partial y^2}.
    \label{laplace_operator}
\end{equation}
It is assumed that the Reynolds number is sufficiently small to neglect inertial forces acting on the fluid. Velocities are non-dimensionalized with respect to $s L^2/\mu$, with $L$ being a characteristic length of the respective flow regime. 
Furthermore, the finite slip boundaries follow the shape given by the pipe geometry. That is, the surface tension is large enough to prevent further curvature of the interface. 

Due to the linearity of the Poisson equation, the resulting velocity field is also linear under consideration of appropriate boundary conditions. Thus, flow fields can be superposed and the result is also a linear solution of the governing partial differential equation. Consider 
\begin{equation}
    \mu \nabla^2 w_1 + \mu \nabla^2 w_2 = \nabla p_1 + \nabla p_2, 
    \label{superposition_1}
\end{equation}
with additivity $f(x_1)+f(x_2)=f(x_1+x_2)$ yielding
\begin{equation}
    \mu \nabla^2 (w_1 + w_2)= \nabla (p_1 + p_2)  ,
    \label{superposition_2}
\end{equation}	
with the superposed velocity field $w=w_1+w_2$ and pressure gradient $p=p_1+p_2$. In this work, available no-shear velocity fields are superposed with suitable solutions of the Poisson equation. With this method, previous no-shear solutions will be extended in such a way that they possess a finite local slip length instead of an infinite one along their interface. 

\subsection{LIS pipe with patterned wall} \label{subsec:math_description_pipe}
\begin{figure}
    \centerline{\includegraphics{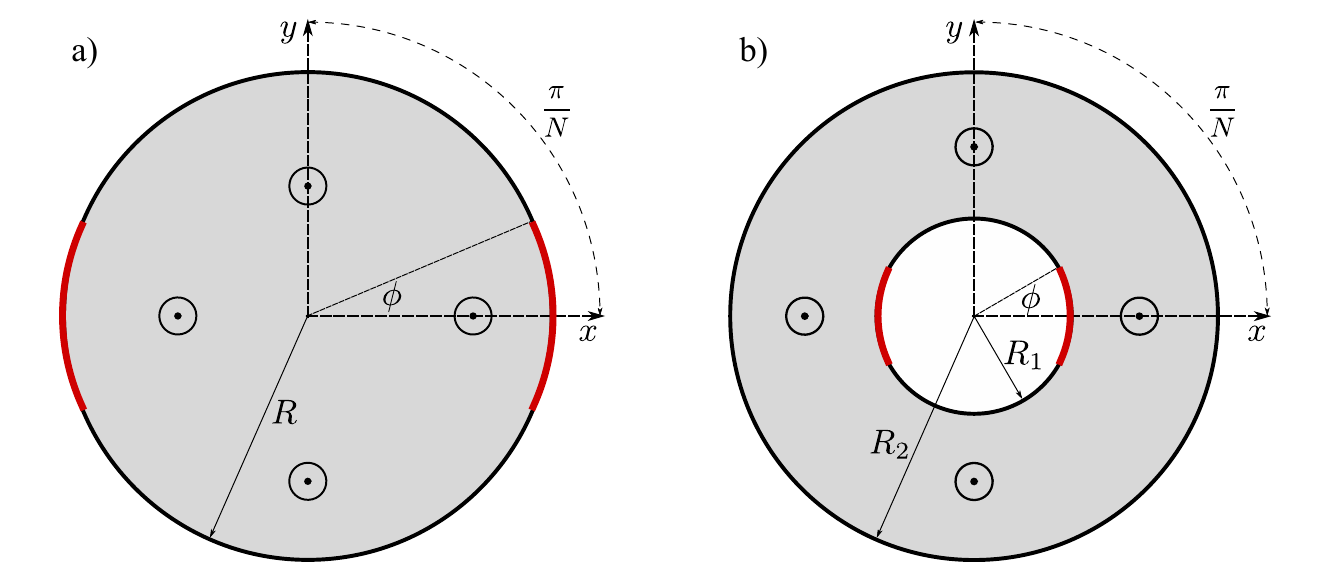}}
    \caption{LIS pipes with mixed boundary conditions, no-shear (or finite constant shear) slits shown as red bold line segments along the otherwise no-slip boundaries. a) Pipe flow case with $N=2$. b) The annular pipe flow with $N=2$ slits on the inner wall. }
    \label{fig:fig_pipe_annular_geometry}
\end{figure}

A flow in a circular tube with $N$ longitudinal rotationally symmetric slits on the outer wall is first considered, as illustrated in figure \ref{fig:fig_pipe_annular_geometry}a, where $N \geq 1$ is an integer. 

\subsubsection{Velocity field - No-shear solution} \label{subsubsec:math_description_pipe_no-shear}
A very important theoretical reference for modelling superhydrophobic or slippery surfaces is \citet{Philip_1972a}. He offers analytic solutions to a variety of mixed value boundary problems which are composed of a mixture of no-shear and no-slip boundaries. One of these solutions describes the aforementioned circular tube with longitudinal no-shear slits on the outer wall, which will be referred to as Philips pipe flow hereafter. The velocity field for an arbitrary dimensionless pipe radius $0 < |\tilde{z}| \leq \tilde{R}$ can be written as \citep{LaugaStone_2003}
\begin{equation}
    w_{\mathrm{pc}}(\tilde{z}) =  \frac{s R_0^2}{\mu} \left(   \frac{1}{4} \left(\tilde{R}^2- |\tilde{z}|^2 \right)  + \frac{1}{N} \tilde{R}^2 \tau(\tilde{z})  \right)
    \label{eq:pipeflow_solution_no-shear}
\end{equation}
where $R_0$ is the dimensionally dependent pipe radius, $\tilde{R} = R/R_0$ the normalized outer pipe radius and $\tilde{z}=z/R_0= \tilde{x} + \mathrm{i} \tilde{y}$ the dimensionless complex coordinate. The index pc indicating the circular Philip solution. It consists of a rotationally axis-symmetric Poisseuille flow superposed by an asymmetric second part with $\tau(\tilde{z})$ governing the no-shear slit influence on the flow, given by
\begin{equation}
    \tau(\tilde{z}) =   \Imag \Bigg[{\cos^{-1}\left({\frac{\cos(\kappa(\tilde{z}))}{\cos\left(\frac{\theta}{2}\right)}}\right)}-\kappa(\tilde{z}) \Bigg]
\end{equation}
and
\begin{equation}
    \kappa(\tilde{z}) = - \frac{\mathrm{i}}{2} \ln\left( \frac{\zeta}{\tilde{R}^N}\right)=- \frac{\mathrm{i}}{2} \ln\left( \frac{\tilde{z}^N}{\tilde{R}^N}\right) = -\frac{\mathrm{i} N}{2} \ln\left(\frac{\tilde{z}}{\tilde{R}}\right) . 
\end{equation}
$\zeta=\tilde{z}^N$ is the transformed coordinate as a result of the conformal mapping performed by Philip and $\Imag$ denotes the imaginary part of an expression. As in section \S\ref{sec:intro}, $\theta = N \phi$ is a measure for the no-shear fraction, with $\phi$ being the slit half angle.

\subsubsection{Velocity field - Finite shear solution} \label{subsubsec:math_description_pipe_finite-shear}
Philips pipe flow solution has perfect slip along the fluid-fluid interfaces, resulting in an infinite local slip length. As mentioned, this assumption corresponds to an ideal state of fluid-fluid interaction, non-existent in reality \citep{bolognesi2014evidence}. To close this gap, one needs to find a solution with finite slip along the slit parts of the boundary. Such a flow field can be described by a superposition ansatz \citep{Schoenecker_2014_Influenceenclosedfluid}
\begin{equation}
    w(\tilde{z}) = A_1  w_{1}(\tilde{z}) + A_2  w_2(\tilde{z}),
\end{equation}
with $w_{1}(\tilde{z})$ being the aforementioned Philip pipe flow solution $w_{\mathrm{pc}}(\tilde{z})$. The underlying assumption of this model is that the shear stress across the fluid-fluid interface is constant. Such a condition has already been used to describe finite-shear models, such as \citet{Schoenecker_2013_singleCav}. Numerical simulations also show that the shear stress is indeed almost constant along the interface \citep{Schoenecker_2014_Influenceenclosedfluid, Higdon_1985_Stokesflowarbitrary}. Exceptions are the regions near the corners of the groove. However, these regions are much smaller than the interface, so the deviation from our assumption is negligible. Furthermore, the approach is closer to reality compared to the assumption of constant slip length along the interface, as the latter inevitably leads to slip length discontinuities at the groove corners. To account for constant shear along the interface, $w_2(\tilde{z})$ is chosen to be Poiseuille flow
\begin{equation}
    w_2(\tilde{z}) = \frac{1}{4} \frac{s_{2} R_0^2}{\mu}  \left( \tilde{R}^2-|\tilde{z}|^2 \right).
\end{equation} 
The superposition ansatz thus gives
\begin{equation}
    w(\tilde{z}) =  A_1 \frac{1}{4} \frac{s_{1}  R_0^2}{\mu} \left(  \left( \tilde{R}^2-|\tilde{z}|^2 \right) + \frac{\tilde{R}^2}{N} \tau(\tilde{z}) \right)  + A_2 \frac{1}{4} \frac{s_{2}  R_0^2}{\mu} \left( \tilde{R}^2-|\tilde{z}|^2 \right) 
\end{equation}
with constants $A_1$ and $A_2$ to be determined. Note that $w_{1}$ and $w_{2}$ are driven by (different) pressure gradients $s_1$ and $s_2$, respectively. Under the condition that the combined flow is driven by a combined pressure gradient 
\begin{equation}
    s = A_1 \ s_1 + A_2 \ s_2,
\end{equation} 
and that the Navier-slip condition applies in the center of the slit at $\tilde{\mathfrak{z}} = \tilde{R} + i 0$
\begin{equation}
    w(\tilde{\mathfrak{z}}) = \lambda \frac{\partial w(\tilde{\mathfrak{z}})}{-\partial \boldsymbol{n}},
\end{equation} 
the constants $A_1$ and $A_2$ are readily determined. The resulting superposed velocity field solution is
\begin{equation}
    w(\tilde{z}) = \frac{s R_0^2}{\mu} \left(  \frac{1}{4}  \left(\tilde{R}^2- |\tilde{z}|^2 \right) + \alpha \frac{\tilde{R}^2}{N}    \tau(\tilde{z})\right) ,
    \label{eq:philip-pipe-flow_finite-shear}
\end{equation} 
and normalized
\begin{equation}
    \tilde{w}(\tilde{z}) = w(\tilde{z}) \left(\frac{s R_0^2}{\mu} \right)^{-1}  =\frac{1}{4}  \left(\tilde{R}^2- |\tilde{z}|^2 \right) + \alpha \frac{\tilde{R}^2}{N}    \tau(\tilde{z}),
    \label{eq:philip-pipe-flow_finite-shear_norm}
\end{equation} 
with
\begin{equation}
    \alpha = \frac{\tilde{\lambda} N}{\tilde{\lambda} N + 2 \tilde{R} \tau(\tilde{\mathfrak{z}}) },
    \label{eq:philip-pipe-flow_alpha}
\end{equation}
where $\tilde{\lambda} = \lambda / R_0$ is the dimensionless local slip length at the slit centre and 
\begin{equation}
    \tau(\tilde{R} + \mathrm{i} 0) = \tau(\tilde{\mathfrak{z}}) = \cosh^{-1}\left(\sec\left(\frac{\theta }{2} \right)  \right). 
\end{equation}
By comparing the superposed flow field with Philips pipe flow, we find that both solutions only differ by the coefficient $\alpha$ in the second term. As $\tilde{\lambda} \rightarrow \infty$, $\alpha$ converges to 1 and $w(\tilde{z}) \rightarrow w_{\mathrm{pc}}(\tilde{z})$, yielding the no-shear solution as it should. $\alpha$ can therefore be interpreted as an imperfection coefficient adjusting the slit influence on the velocity field depending on a potentially finite local slip length. 

\subsubsection{Effective slip length} 
\label{subsubsec:math_description_pipe_eff-slip}
With the flow field in dependence on a local slip length at hand, it is now possible to calculate the effective slip length. It represents the averaged influence of the slip boundary parts on the velocity field and involves equating the total volume flux $\dot{V}$ caused by a given pressure gradient to that of a suitable comparison flow $(0,0,w^*(x,y))$ with the same pressure gradient, containing a no-slip wall at $|z| = R$. The effective slip length is the value of $\lambda_{\mathrm{eff}}$, for which $\dot{V}^* = \dot{V}(\lambda_{\mathrm{eff}})$.  In our case, however, it is not necessary to calculate the volume flow. Instead we consider a pressure-driven pipe flow with some constant effective slip length $\lambda_{\mathrm{eff}}$ at the outer wall. This will provide a simple formula that can be used to calculate the effective slip length.  First, a general axissymmetrical pipe flow solution is given by
\begin{equation}
    w^* = - \frac{s}{4 \mu} |z|^2 + c_1 \ln(|z|) + c_2,
    \label{eq:pipe_general-solution}
\end{equation} 
subject to 
\refstepcounter{equation}
$$
    \frac{\partial w^*(|z|=0)}{\partial |z|} = 0, \quad\quad w^*(|z|=R) = \lambda_{\mathrm{eff}} \frac{\partial w^*(|z|=R)}{- \partial |z|},
    \eqno{(\theequation \textit{a,b})}
$$
yields
\begin{equation}
    w^* = \frac{1}{4} \frac{s}{\mu} (R^2 - |z|^2) + \frac{1}{2} \frac{s}{\mu} R \lambda_{\mathrm{eff}},
    \label{eq: effective-slip-length_velocity-field_pipe-flow}
\end{equation}  
which corresponds to a pipe flow with constant slip $\lambda_{\mathrm{eff}}$ at the outer wall. This can be interpreted as the effective slip length.  From equation \ref{eq: effective-slip-length_velocity-field_pipe-flow}, an expression for $\lambda_{\mathrm{eff}}$ at $|z|=R$ is easily determined to be
\begin{equation}
    \lambda_{\mathrm{eff}} =  \frac{\mu}{s} \frac{2}{R} w^*(R). 
    \label{eq:pipe-flow_lambda-eff_formula}
\end{equation}  
As mentioned above, all groove influence is abstracted to an average effect on the given wall. So all left to do is to average equation \ref{eq:philip-pipe-flow_finite-shear} along the boundary
\begin{equation}
    w^*(R) = \frac{1}{2 \upi} \int_{0}^{2 \upi} w(R) d \varphi= \alpha \frac{s}{\mu}\frac{R^2}{N}   \frac{1}{2 \upi} \int_{0}^{2 \upi} \tau(\tilde{z}) \ \mathrm{d} \varphi, 
\end{equation} 
where $\varphi$ is the coordinate angle. Solving the above integral involves using the integral relation derived by \citet{Philip_1972b}, stating that the averaged slip influence is given by
\begin{equation}
    \frac{1}{2 \upi} \int_{0}^{2 \upi} \tau(\tilde{z}) \ \mathrm{d} \varphi = \bar{\tau}(\tilde{z}) = \ln\left(\sec\left(\frac{\theta}{2} \right)  \right). 
\end{equation} 
The resulting normalized effective slip length $\tilde{\lambda}_{\mathrm{eff}} = \lambda_{\mathrm{eff}}/R $ is
\begin{equation}
    \tilde{\lambda}_{\mathrm{eff}} = \alpha \frac{2}{N} \ln\left(\sec\left(\frac{\theta}{2} \right)  \right). 
    \label{eq:philip-pipe-flow_effective-slip-norm}
\end{equation} 
Equation \ref{eq:philip-pipe-flow_effective-slip-norm} again transitions into Philip solution if the local slip length diverges. However, there is a more straightforward way to calculate the effective slip length. Equation \ref{eq:pipe-flow_lambda-eff_formula} shows, that the Poisseuille part is not contributing to the effective slip length, since it is zero at $|z|=R$, while the second part is non-zero on the boundary and contributes with the factor of $\alpha$. The normalized slip length can therefore be directly determined by 
\begin{equation}
    \tilde{\lambda}_{\mathrm{eff}} = \alpha \ \tilde{\lambda}_{\mathrm{eff},P},
\end{equation} 
with $\tilde{\lambda}_{\mathrm{eff},P}$ being the effective slip length of Philip's \citep{Philip_1972a} no-shear solution. 

\subsubsection{Volume flow}\label{subsubsec:math_description_pipe_vol-flux}
Another useful result is the total volume flux generated by the superposed flow of eq. \ref{eq:philip-pipe-flow_finite-shear}. For this purpose we consider the volume flow to be given by
\begin{equation}
    \dot{V} = 2 \upi \int_{0}^{R} w(x,y) \ r \ \mathrm{d}r = 2 \frac{s \upi}{\mu}  \int_{0}^{R} \left(  \frac{1}{4}  \left(R^2-|z|^2 \right) +  \alpha \frac{R^2}{N} \bar{\tau}(\tilde{z})\right) \ r \  \mathrm{d}r ,
\end{equation} 
where we again use the integral identity of \citet{Philip_1972b}. The associated total volume flux therefore is
\begin{equation}
    \dot{V} = \frac{s \pi}{\mu} \left(\frac{1}{8} R^4 + \alpha \frac{R^4}{N} \ln\left(\sec\left(\frac{\theta}{2} \right)  \right)  \right).
\end{equation} 
As in \ref{eq:philip-pipe-flow_finite-shear}, with $\tilde{\lambda} \rightarrow \infty$ the volume flux transitions into the Philip solution \citep{Philip_1972b}, representing the ideal no-shear limit. With the comparison volume flux of a Poiseuille flow with no-slip at the outer wall being
\begin{equation}
    \dot{V}_{\mathrm{ns}} = \frac{1}{8} \frac{s \pi}{\mu} R^4,
\end{equation} 
the effective slip length can alternatively be calculated with
\begin{equation}
    \lambda_{\mathrm{eff}} = \frac{R}{4} \left(\frac{\dot{V}}{\dot{V}_{\mathrm{ns}}} -1 \right),
\end{equation} 
which also results in equation \ref{eq:philip-pipe-flow_effective-slip-norm}. 

\subsection {Annular LIS pipe with patterned inner wall} 
\label{subsec:math_description_annulus}
A second flow regime of great interest is an annular slippery pipe with $N$ rotationally symmetric slits on the inner boundary wall, as illustrated in figure \ref{fig:fig_pipe_annular_geometry}b. 

\subsubsection{Velocity field - No-shear solution}
\label{subsubsec:math_description_annulus_no-shear}
\citet{Crowdy_2021_theoStud} provides an analytic flow field solution for annular superhydrophobic pipes of radius $\tilde{R}_1 \leq |\tilde{z}| \leq \tilde{R}_2$ with $\tilde{R}_2=1$ and $N \geq 1$ no-shear boundary slits on the inner wall. The velocity field is given to be 
\begin{equation}
    w_{\mathrm{ca}}(\tilde{z}) = \frac{s R_0^2}{\mu} \left( \frac{1}{4} \left( 1-|\tilde{z}|^2 \right)  +  \frac{1}{2} \Real(H(\zeta)) \right),
    \label{eq:annular-pipeflow_solution_no-shear} 
\end{equation}
where the subscript ca indicates the Crowdy annular flow field \citep{Crowdy_2021_theoStud}. Like equation \ref{eq:pipeflow_solution_no-shear}, this solution consists of a rotationally symmetric Poiseuille flow and a superposed asymmetric second term with the real part $\Real$ of an analytic function $H(\zeta)$. This function represents the slit influence on the flow field and is 
\begin{equation}
    H(\zeta) = \frac{1}{N} \int_{-1}^{\zeta}  \Bigg[\tilde{R}_1^2-M  \left( \frac{P\left( \frac{\zeta'}{q},q \right) P\left( \frac{\zeta'}{q},q\right)  }{P\left( \frac{\zeta'}{a},q\right)P\left( \frac{\zeta'}{\bar{a}},q\right)}\right)^{1/2} \Bigg]  \frac{\mathrm{d} \zeta'}{\zeta},
\end{equation}
where $a=\tilde{R}_1^N e^{i\theta}$ and $q=\tilde{R}_1^N$. The angle $\theta = \phi N$ is defined in the same way as for Philip's pipe solution. The variable $\zeta = z^N$ is, as before, the transformed complex coordinate. $M$ is a scaling factor determined by imposing $w_{\mathrm{ca}}(x,y)=0$ on the no-slip portions of the inner wall 
\begin{equation}
    M = \frac{\frac{(1-\tilde{R}_1^2)}{4}+\frac{1}{2} \tilde{R}_1^2  \ln(\tilde{R}_1)}{S},
    \label{eq:annular-pipeflow_m}
\end{equation}
with 
\begin{equation}
    S =  \frac{1}{2 N} \int_{-1}^{-q}  \left( \frac{P\left( \frac{\zeta}{q},q\right)P\left( \frac{\zeta}{q},q\right) }{P\left( \frac{\zeta}{a},q\right)P\left( \frac{\zeta}{\bar{a}},q\right)}\right)^{1/2} \frac{d\zeta}{\zeta}.
    \label{eq:annular-pipeflow_s}
\end{equation}
$P(\zeta,q)$ is called the prime function for the concentric annulus and is defined as a convergent infinite product for any $\zeta \neq 0$ 
\begin{equation}
    P(\zeta,q) = (1-\zeta) \prod_{n=1}^{\infty} (1-q^{2n} \cdot \zeta)(1-q^{2n}/\zeta), \quad 0 \leq q < 1. 
    \label{eq:prime-function}
\end{equation}
For further information on prime functions, see \citet{Crowdy_2020_BOOK}. 

\subsubsection{Velocity field - Finite shear solution}
\label{subsubsec:math_description_annulus_finite-shear}
As for Philip's pipe flow, we seek a solution that does not necessarily impose an infinite local slip length on the slit portions of the boundary, but rather has a potentially finite local slip length. As before, constant shear stress along the fluid-fluid interface is assumed. We again choose a superposition approach
\begin{equation}
    w(\tilde{z}) = B_1 \ w_{1}(\tilde{z}) + B_2 \ w_2(\tilde{z}),
\end{equation}
with $w_{1}(\tilde{z})$ being the no-shear solution for the annulus provided by \citet{Crowdy_2021_theoStud}. $w_2(\tilde{z})$ is chosen to be the Poiseuille solution for an annular domain, with $w_2(\tilde{z})=0$ at $|\tilde{z}|=\tilde{R}_1$ and $|\tilde{z}|=1$, resulting in 
\begin{equation}
    w_{2}(\tilde{z}) = \frac{1}{4} \frac{s_{2} R_0^2}{\mu} \left( \left(1-|\tilde{z}|^2 \right) - \left( 1-\tilde{R}^2_1\right) \frac{\ln(|\tilde{z}|)}{\ln(\tilde{R}_1)}    \right) . 
    \label{eq:annular-pipeflow_solution_no-slip}
\end{equation} 
The superposition ansatz thus gives
\begin{eqnarray}
    w(\tilde{z}) &=& B_1 \frac{s_1 R_0^2}{\mu} \left(  \frac{1}{4} (1-|\tilde{z}|^2) + \frac{1}{2}  \Real(H(\zeta)) \right) \nonumber \\
    && + B_2 \frac{1}{4} \frac{s_{2} R_0^2}{\mu} \left( \left(1-|\tilde{z}|^2 \right) - \left( 1-\tilde{R}^2_1\right) \frac{\ln(|\tilde{z}|)}{\ln(\tilde{R}_1)}    \right),
\end{eqnarray}
with constants $B_1$ and $B_2$ again to be determined. $w_{1}$ and $w_{2}$, as before, are driven by the pressure gradients $s_1$ and $s_2$, respectively. Under the condition that the combined flow is driven by a combined pressure gradient 
\begin{equation}
    s = B_1 \ s_1 + B_2 \ s_2,
\end{equation} 
and with the Navier-slip condition imposed at the center of the slit $\tilde{\mathfrak{z}} = \tilde{R_1} + i 0$
\begin{equation}
    w(\tilde{\mathfrak{z}}) = \lambda \frac{\partial w(\tilde{\mathfrak{z}})}{\partial \boldsymbol{n}}, 
\end{equation} 
the constants $B_1$ and $B_2$ are determined, yielding the corresponding superposed flow field
\begin{equation}
    w(\tilde{z}) =  \frac{1}{4} \frac{s R_0^2}{\mu} \left(  (1-|\tilde{z}|^2) + 2 \beta_1  \Real(H(\zeta))  -  \beta_2  \left( 1-\tilde{R}^2_1\right) \frac{\ln(|\tilde{z}|)}{\ln(\tilde{R}_1)}    \right) ,
    \label{eq:annular-pipeflow_solution_finite-slip}
\end{equation} 
with
\begin{equation}
    \beta_1 = \frac{\tilde{\lambda} \left(\tilde{R}_1^2+2 \tilde{R}_1^2 \ln\left(\frac{1}{\tilde{R}_1} \right) -1  \right) }{ \Big[ \tilde{R}_1^2 \tilde{\lambda}- \tilde{\lambda} + \tilde{R}_1 \ln\left(\frac{1}{\tilde{R}_1} \right) \bigg(\tilde{R}_1^2-2 \Real(H(q))+2 \tilde{\lambda}\tilde{R}_1 -1 \bigg)  \Big] } ,
    \label{eq:annular-pipeflow_beta1}
\end{equation} 
and 
\begin{equation}
    \beta_2 = (1- \beta_1)= \frac{\tilde{R}_1 \ln\left(\frac{1}{\tilde{R}_1} \right) \bigg(\tilde{R}_1^2-2 \Real(H(q))-1 \bigg) }{\Big[ \tilde{R}_1^2 \tilde{\lambda}- \tilde{\lambda} + \tilde{R}_1 \ln\left(\frac{1}{\tilde{R}_1} \right) \bigg(\tilde{R}_1^2-2 \Real(H(q))+2 \tilde{\lambda}\tilde{R}_1 -1 \bigg)  \Big] },
    \label{eq:annular-pipeflow_beta2}
\end{equation} 
where $\tilde{\lambda}= \lambda/R_0$ is the normalized local slip length and $H(q)$ is the value of the analytic function $H(\zeta)$ evaluated at the center of the slit boundary portion, at coordinate $\tilde{\mathfrak{z}}$, corresponding to $\zeta(\tilde{\mathfrak{z}})=q$. 
Analysing the behaviour of equation \ref{eq:annular-pipeflow_solution_finite-slip} in dependence of coefficients $\beta_1$ and $\beta_2$ shows that for $\tilde{\lambda} \rightarrow \infty$, $\beta_1 \rightarrow 1$ and $\beta_2 \rightarrow 0$ delivering the no-shear solution of equation \ref{eq:annular-pipeflow_solution_no-shear}. Therefore both coefficients can be referred to as weighting coefficients,  steering the influence of the no-shear \ref{eq:annular-pipeflow_solution_no-shear} and no-slip solution \ref{eq:annular-pipeflow_solution_no-slip} on the superposed flow field. The normalized velocity field is easily calculated to be
\begin{equation}
    \tilde{w}(\tilde{z}) =   \frac{1}{4} \left( (1-|\tilde{z}|^2) + 2 \beta_1  \Real(H(\zeta))  -  \beta_2  \left( 1-\tilde{R}^2_1\right) \frac{\ln(|\tilde{z}|)}{\ln(\tilde{R}_1)}    \right).
    \label{eq:annular-pipeflow_solution_finite-slip_norm}
\end{equation} 

\subsubsection{Effective slip length}
\label{subsubsec:math_description_annulus_eff-slip}
As with the pipe flow in section \S\ref{subsubsec:math_description_pipe_eff-slip}, a suitable comparison flow is needed to determine the effective slip length for the superposed annular flow above. For that, the general axissymmetrical solution of eq. \ref{eq:pipe_general-solution} is solved again, with velocity normalized with respect to $s R_0^2/ \mu$ and length scales with $R_0$. This solution is subject to no-slip on the wall at $|\tilde{z}| = \tilde{R}_2 = 1$ and a Navier-slip boundary condition at $|\tilde{z}| = \tilde{R}_1$, so 
$$
    \refstepcounter{equation}
    \tilde{w}^*(|\tilde{z}|=\tilde{R}_2) = 0, \quad\quad \tilde{w}^*(|\tilde{z}|=\tilde{R}_1) = \tilde{\lambda} \ \frac{\partial \tilde{w}^*(|\tilde{z}|=\tilde{R}_1)}{\partial |\tilde{z}|}.
    \eqno{(\theequation \textit{a,b})}
$$
Implementing both boundary conditions yields an annular axissymmetric solution with a constant slip length at $|\tilde{z}| = \tilde{R}_1$, which can be interpreted as an effective slip along that boundary $\tilde{\lambda} \equiv \tilde{\lambda}_{\mathrm{eff}}$. The solution for the flow field is found to be
\begin{equation}
    \tilde{w}^* =  \frac{1-|\tilde{z}|^2}{4} - \tilde{R}_1 \ln(|\tilde{z}|) \frac{\tilde{R}_1^2-2 \tilde{R}_1 \tilde{\lambda}_{\mathrm{eff}}-1}{4 (\tilde{\lambda}_{\mathrm{eff}} - \tilde{R}_1 \ln(\tilde{R}_1))}.  
\end{equation} 
Evaluating this solution at the inner slip wall $|\tilde{z}| = \tilde{R}_1$ gives
\begin{equation}
    \tilde{w}^*(|\tilde{z}|=\tilde{R}_1) = \frac{1-\tilde{R}_1^2}{4} - \tilde{R}_1 \ln(\tilde{R}_1) \frac{\tilde{R}_1^2-2 \tilde{R}_1 \tilde{\lambda}_{\mathrm{eff}}-1}{4 (\tilde{\lambda}_{\mathrm{eff}} - \tilde{R}_1 \ln(\tilde{R}_1))}.
\end{equation} 
A rearrangement provides a formula for the effective slip length on the inner wall of the annulus $\tilde{R}_1 \leq |\tilde{z}| \leq 1$ as a function of the normalized velocity $\tilde{w}^*$ on that very wall
\begin{equation}
    \tilde{\lambda}_{\mathrm{eff}} = - \frac{4 \tilde{R}_1 \ln(\tilde{R}_1) \tilde{w}^*}{(1- \tilde{R}_1^2 - 4 \tilde{w}^* + 2 \tilde{R}_1^2 \ln(\tilde{R}_1) )}. 
    \label{eq:annular-flow_effective-slip_formula}
\end{equation} 
As mentioned earlier, the effective slip length abstracts the cumulative groove influence on the flow field as an average effect along the respective wall. Equation \ref{eq:annular-pipeflow_solution_finite-slip_norm}, however, is based on mixed-boundary conditions along the groove containing wall. It is easy to see that the velocity is not constant, since it is zero at the no-slip portions and non-zero along the slits. Therefore, it is necessary to average the velocity along $|\tilde{z}| = \tilde{R}_1$ and thus indirectly the groove effect. A closer look at the equations \ref{eq:annular-pipeflow_solution_no-shear} and \ref{eq:annular-pipeflow_solution_finite-slip_norm} reveals that the only rotationally asymmetric terms are those containing the real part of the analytic function $H(\zeta)$. The no-shear and finite shear solutions are thus closely related with respect to their asymmetric behaviour. Therefore, it makes sense to first consider the effective slip length of the former. The finite slip solution can then be considered as a simple extension of it. Reconstructing the effective slip length of the no-shear solution derived by \citet{Crowdy_2021_theoStud}, we must first identify the average of $\Real(H(\zeta))$ at $|\tilde{z}|=\tilde{R}_1$ along the inner wall
\begin{equation}
    \Real(H(\zeta))_{\mathrm{avg}} = \frac{1}{2 \pi} \int_{0}^{2 \pi} \Real(H(\zeta)) \mathrm{d} \varphi. 
\end{equation} 
Although not explicitly given in \citet{Crowdy_2021_theoStud}, it is readily determined to be
\begin{equation}
    \Real(H(\zeta))_{\mathrm{avg}} = \tilde{R}_1^2 \ln(\tilde{R}_1) - M I_{-1} \ln(\tilde{R}_1),
\end{equation} 
with the aforementioned scaling factor $M$ from equation \ref{eq:annular-pipeflow_m} and 
\begin{equation}
    I_{-1}=  \frac{1}{2 \upi i}  \oint_{C} \left( \frac{P\left( \frac{\zeta}{q},q\right)P\left( \frac{\zeta}{q},q\right) }{P\left( \frac{\zeta}{a},q\right)P\left( \frac{\zeta}{\bar{a}},q\right)}\right)^{1/2} \frac{\mathrm{d}  \zeta}{\zeta},
\end{equation}  
where $C$ is any closed circle inside the annulus $q < |\zeta| < 1$ enclosing the origin \citep{Crowdy_2021_theoStud}. Accordingly, the averaged velocity for the no-shear solution along the inner wall is given by 
\begin{equation}
    \tilde{w}^*_{\mathrm{ca}}(|\tilde{z}|=\tilde{R}_1) = \frac{1}{4} (1-\tilde{R}_1^2) + \frac{1}{2} (\tilde{R}_1^2 \ln(\tilde{R}_1) - M I_{-1} \ln(\tilde{R}_1)),
\end{equation} 
and the associated normalized effective slip length for the no-shear case is 
\begin{equation}
    \tilde{\lambda}_{\mathrm{eff},\mathrm{ca}} = \frac{\lambda_{\mathrm{eff},\mathrm{ca}}}{R_0}=  \tilde{R}_1 \ln(\tilde{R}_1) - 2 \tilde{R}_1 \frac{S}{I_{-1}},
    \label{eq:annular-flow_effective_slip_no-shear}
\end{equation} 
with $S$ from equation \ref{eq:annular-pipeflow_s}, as is given in \citet{Crowdy_2021_theoStud}. Determining the effective slip for the superposed annular flow is performed similarly. Equation \ref{eq:annular-pipeflow_solution_finite-slip_norm} shows, that the annular flow multiplied by $\beta_2$ (third term in brackets) does not contribute to $\lambda_{\mathrm{eff}}$, since it is zero at the inner wall. Therefore
\begin{equation}
    \tilde{w}^*= \beta_1 \tilde{w}^*_{\mathrm{ca}} =\beta_1 \frac{1}{4} (1-\tilde{R}_1^2) + \beta_1 \frac{1}{2} (\tilde{R}_1^2 \ln(\tilde{R}_1) - M I_{-1} \ln(\tilde{R}_1))
\end{equation} 
at $|\tilde{z}| = \tilde{R}_1$. From that, with equation \ref{eq:annular-flow_effective-slip_formula}, the effective slip length for the finite shear case is easily calculated to be
\begin{equation}
    \tilde{\lambda}_{\mathrm{eff}} = \frac{\lambda_{\mathrm{eff}}}{R_0} = \tilde{R}_1 \ln(\tilde{R}_1) \beta_1 \frac{I_{-1} \ln(\tilde{R}_1) - 2 S}{I_{-1} \ln(\tilde{R}_1) \beta_1 - 2 S (\beta_1-1)}. 
    \label{eq:annular-flow_effective_slip_combined}
\end{equation}  
For $\beta_1 \rightarrow 1$, representing an infinite local slip length on the slit boundary parts, equation \ref{eq:annular-flow_effective_slip_combined} converges to the effective slip length of the no-shear solution in equation \ref{eq:annular-flow_effective_slip_no-shear}, as expected. 

\section {Coupled flow - Pipe within a pipe} 
\label{sec:coupling}
\begin{figure}      
    \centerline{\includegraphics{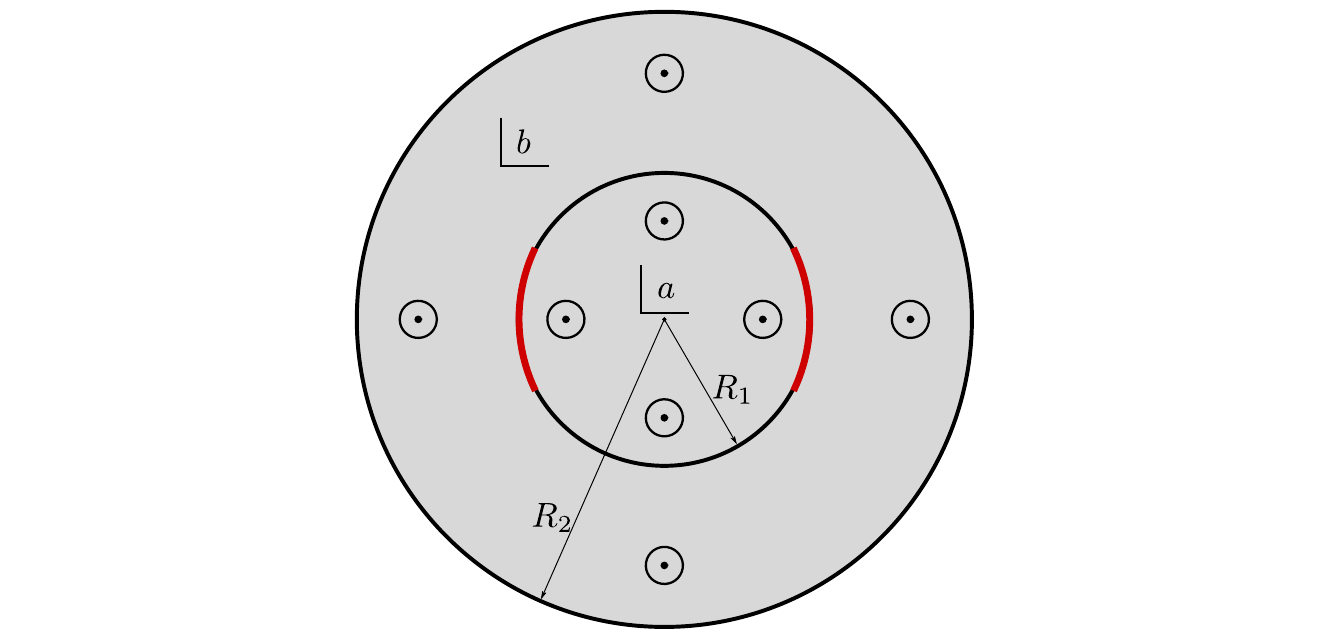}}
    \caption{Fluid domain connection - Pipe flow (domain a) and annular flow (domain b) are connected via two rotationally symmetric shear slits on the inner wall, illustrated as bold red slits.}
    \label{fig:connection_pipe_annular_inf-plates_geometry}
\end{figure}

Section \S\ref{sec:math_description} provides, among other things, analytic mathematical expressions for the pressure-driven flow field of a circular pipe  (section \S\ref{subsec:math_description_pipe}) and an annular pipe flow (section \S\ref{subsec:math_description_annulus}), both containing an periodic array of longitudinal grooves along one boundary. Both solutions presented stand for themselves. However, they do not  assume perfect slip but rather a constant finite shear-stress along these grooves, contrary to most literature (\citet{Lee_2016_Superhydrophobicdragreduction, LaugaStone_2003, schnitzer2019stokes,Crowdy_2021_theoStud, sbragaglia2007note, Crowdy_2016_Analyticalformulaelongitudinal, yariv2018pressure}). This is represented by a finite local slip length at the fluid-fluid interface, which is to be determined depending on the respective application. 
In the case of LIS, this could be longitudinal grooves or other roughness features along the surface filled with a second immiscible fluid, usually air or oil. However, the practical application of such surfaces is challenging, as the trapped fluid can be pushed out of the microstructure. The interfacial collapse then leads to a Cassie-Wenzel transition. This ruins the desired effects of these surfaces, such as drag reduction. One way to prevent this is to use bottomless surfaces, where the downward walls inside the microstructures have been removed. This provides the operator with a greater degree of control. External pressure control, for example, can prevent the interface from collapsing. A possible configuration of such a surface is a pipe within a pipe, as illustrated in figure \ref{fig:connection_pipe_annular_inf-plates_geometry}. Such a domain is composed of a pipe flow (a) and an annular flow (b), which are connected along certain boundary areas, illustrated as red circular arcs in figure \ref{fig:connection_pipe_annular_inf-plates_geometry}. This makes it possible, for example, to increase the interfacial stability by controlling both flow fields accordingly. Obviously, the pipe within a pipe geometry is composed of the solutions for the pipe flow and the annular flow derived in section \ref{sec:math_description}, both of which have a local slip length as an unknown to be determined. These local slip lengths can be readily calculated as parameters depending on the corresponding linked flow field, so $\lambda_a = f(w_b)$ and vice versa. To emphasize that $\lambda_{a,b}$ follow from the connection of the two flow fields, it will be referred to here as $\Lambda_{a,b}$.   

Accordingly, a pipe within a pipe is considered, connected through an periodic array of rotationally symmetric longitudinal slits on the outer wall of the inner circle. The no-slip wall portions of this boundary are assumed to be infinitely thin. Both domains are assumed to be pressure-driven. The outer radius of the governing equation of the inner pipe flow (eq. \ref{eq:philip-pipe-flow_finite-shear}) is scaled to be of radius $\tilde{R}_1$, which corresponds to the inner radius of the annular flow solution. It is given by
\begin{equation} 
    w_a(\tilde{z}) = \frac{s_a R_0^2}{\mu_a} \left(  \frac{1}{4}  \left(\tilde{R}_1^2- |\tilde{z}|^2\right) + \alpha(\tilde{\Lambda}_a) \frac{\tilde{R}_1^2}{N}    \tau(\tilde{z})\right) ,
    \label{eq:philip-pipe-flow_finite-slip_regime-a}
\end{equation} 
with $s_a, \mu_a$ being the negative pressure gradient and viscosity within fluid domain $a$. The imperfection coefficient $\alpha$ is now  defined in dependence on the normalized local connection slip length $\tilde{\Lambda}_a = \Lambda_a/R_0$ for the inner pipe. $R_0$ is set to be the dimensional radius of the outer pipe. The outer pipe flow is governed by the annular pipe flow solution of eq. \ref{eq:annular-pipeflow_solution_finite-slip}  
\begin{equation} 
    w_b(\tilde{z}) =  \frac{1}{4} \frac{s_b R_0^2}{\mu_b} \left(     (1-|\tilde{z}|^2) + 2 \beta_1(\tilde{\Lambda}_b) \Real(H(\zeta))       -\beta_2(\tilde{\Lambda}_b) \left( 1-\tilde{R}^2_1\right) \frac{\ln(|\tilde{z}|)}{\ln(\tilde{R}_1)}     \right) ,
    \label{eq:annular-pipeflow_solution_finite-slip_regime-b}
\end{equation} 
with $s_b, \mu_b$ of domain $b$ and $\beta_1, \beta_2$ depending on the normalized local connection slip length $\tilde{\Lambda}_b = \Lambda_b/R_0$ for the outer pipe flow. A coupling of both flows has two unknowns, $\tilde{\Lambda}_a (w_b)$ and $\tilde{\Lambda}_b (w_a)$. Two connection conditions are consequently needed. Specifically, both the velocity and shear stress at a single point on the interface must be equal, so  
$$
    \refstepcounter{equation}
    w_a(\tilde{\mathfrak{z}}) = w_b(\tilde{\mathfrak{z}}), \quad\quad \mu_a \frac{\partial w_a(\tilde{\mathfrak{z}})}{- \partial \boldsymbol{n}} = - \mu_b \frac{\partial w_b(\tilde{\mathfrak{z}})}{\partial \boldsymbol{n}},
    \label{eq:pipe-within-pipe_connection-cond_1-2}
    \eqno{(\theequation \textit{a,b})}
$$
evaluated in the centre of the groove at $\tilde{\mathfrak{z}} = \tilde{R}_1 + i  0$. Solving the system of equations (eqs. \ref{eq:pipe-within-pipe_connection-cond_1-2}a,b) yields expressions for the local slip lengths
$$
    \refstepcounter{equation}
    \tilde{\Lambda}_a(w_b) =  \mu_a \Omega_a, \quad\quad \tilde{\Lambda}_b(w_a) =  \mu_b \Omega_b,
    \label{eq:pipe-within-pipe_coupled_lambda_a,b}
    \eqno{(\theequation \textit{a,b})}
$$
where $\Omega_a, \Omega_b$ are the connection coefficients of fluid domain $a$ and $b$, respectively. They are given by
\begin{equation}
    \Omega_a = - \frac{2 \tilde{R}_1  \cosh^{-1}\left(\sec\left( \frac{\theta}{2} \right)  \right) \bigg(\tilde{R}_1^2-2 \Real(H(q))-1 \bigg) \bigg( (\tilde{R}_1^2-1) s_b + 2 \tilde{R}_1^2 \ln\left(\frac{1}{\tilde{R}_1} \right) (s_b-s_a) \bigg) }{\bigg(\tilde{R}_1^2 + 2 \tilde{R}_1^2 \ln \left(\frac{1}{\tilde{R}_1} \right) -1  \bigg) \bigg(N s_b \mu_a (\tilde{R}_1^2-2 \Real(H(q))-1) + 4 \tilde{R}_1^2 s_a \mu_b  \cosh^{-1}\left(\sec\left( \frac{\theta}{2} \right)  \right)   \bigg)},
    \label{eq:pipe-within-pipe_connection-coeff_a}
\end{equation} 
and $\Omega_a = - \Omega_b $. Both coefficients are only dependent on already known quantities and can be easily evaluated. The local connection slip lengths are in proportion to the viscosities of both fluid domains. 

Although unequal pressure gradients $s_a,s_b$ are not explicitly excluded in equation \ref{eq:pipe-within-pipe_connection-coeff_a}, the original assumption to neglect additional flow induced interface curvature is no longer valid, if $s_a \gg s_b$ or $s_a \ll s_b$. However, these equations can be used to consider cases where the local pressure difference $\Delta s = |s_a - s_b|$ at the fluid-fluid interface is less than the Laplace pressure.  
In the special case $s_a=s_b$, the connection coefficients can be simplified to 
\begin{equation}
    \Omega_{a,s} = - \frac{2 \tilde{R}_1  \cosh^{-1}\left(\sec\left( \frac{\theta}{2} \right)  \right) \bigg(\tilde{R}_1^2-2 \Real(H(q))-1 \bigg) \bigg( \tilde{R}_1^2-1 \bigg) }{\bigg(\tilde{R}_1^2 + 2 \tilde{R}_1^2 \ln \left(\frac{1}{\tilde{R}_1} \right) -1  \bigg) \bigg(N \mu_a (\tilde{R}_1^2-2 \Real(H(q))-1) + 4 \tilde{R}_1^2 \mu_b  \cosh^{-1}\left(\sec\left( \frac{\theta}{2} \right)  \right)   \bigg)}, 
    \label{eq:pipe-within-pipe_connection-coeff_s_a}
\end{equation} 
where, again, $\Omega_{a,s} = - \Omega_{b,s}$. The subscript $s$ indicates the same pressure gradient in both flow domains. With both connection coefficients  at hand, the normalized local connection slip length for both flow domains can easily be determined, using equations \ref{eq:pipe-within-pipe_coupled_lambda_a,b}$a,b$. An interesting special case occurs when the same fluid flows in the inner and outer pipe, again with $s_a=s_b$. The coefficients further simplify and yield
\begin{equation}
    \tilde{\Lambda}_{a,s, \mu} = - \frac{2 \tilde{R}_1  \cosh^{-1}\left(\sec\left( \frac{\theta}{2} \right)  \right) \bigg(\tilde{R}_1^2-2 \Real(H(q))-1 \bigg) \bigg( \tilde{R}_1^2-1 \bigg) }{\bigg(\tilde{R}_1^2 + 2 \tilde{R}_1^2 \ln \left(\frac{1}{\tilde{R}_1} \right) -1  \bigg) \bigg(N (\tilde{R}_1^2-2 \Real(H(q))-1) + 4 \tilde{R}_1^2 \cosh^{-1}\left(\sec\left( \frac{\theta}{2} \right)  \right)   \bigg)}, 
\end{equation} 
with $\tilde{\Lambda}_{a,s, \mu} = - \tilde{\Lambda}_{b,s, \mu}$. Index $\mu$ means that $\mu_a = \mu_b$. 

\section {Results and discussion}
\label{sec:results-discussion}
Section \S\ref{sec:math_description} and section \S\ref{sec:coupling} provide mathematical expressions for the flow field and the effective slip length for pipe and annular flow along rotationally symmetric longitudinal grooves as well as their connection at the slit boundary parts to model a pipe-within-pipe geometry. A flow through aforementioned geometries is thus fully described as a function of the pipe radii, number of grooves and a maximum local (connection) slip length at the slit center $\tilde{\mathfrak{z}}$.  

\subsection {Imperfection and weighting coefficients} 
\label{subsec:results-discussion_coefficients}
\begin{figure}
    \centerline{\includegraphics{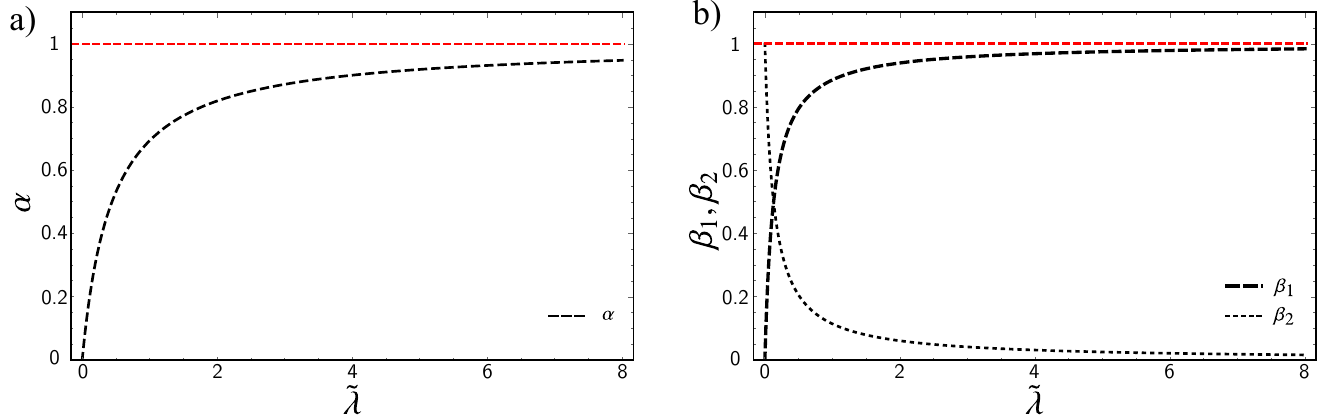}}
    \caption{Progression of imperfection and weighting coefficients with increasing local slip length $\tilde{\lambda}$ for $\theta=\pi/2$, $R_1=0.5$ and $N=2$. a) Imperfection coefficient $\alpha$. b) Weighting coefficients $\beta_1$ and $\beta_2$.   }
    \label{fig:alpha_beta1_beta2_plots}
\end{figure}

Contrary to most literature, the solutions derived in this work do not necessarily have an infinite local slip length at the slit boundary parts, but a potentially finite one. For $\theta=\pi/2$, radius $R_1=0.5$ and two slits, figure \ref{fig:alpha_beta1_beta2_plots} illustrates the progression of the imperfection coefficient $\alpha$ of the superposed pipe flow (eq. \ref{eq:philip-pipe-flow_finite-shear}) and the weighting coefficients $\beta_1,\beta_2$ of the superposed annular flow (eq. \ref{eq:annular-pipeflow_solution_finite-slip}) with increasing local slip length. We consider only $\tilde{\lambda} \geq 0$. The dashed horizontal red line indicates the coefficient limit value of one. 

With no local slip length at the boundary, $\alpha \rightarrow 0$, accordingly the second part of the superposed pipe flow determining the groove influence vanishes, leaving behind a Poiseuille flow with no slip at $|z|=R$. In contrast, the flow field at infinite slip length corresponds to the no-shear solution of \citet{Philip_1972a}, since $\alpha \rightarrow 1$. 
For the superposed annular flow, $\beta_1$ and $\beta_2$ weight the influence of the no-shear solution provided by \citet{Crowdy_2021_theoStud} and the superposed annular Poiseuille flow, respectively. For $\tilde{\lambda} = 0$, $\beta_1$ converges to zero and $\beta_2$ to one, accordingly eq.  \ref{eq:annular-pipeflow_solution_finite-slip} transitions into the annular Poiseuille flow with no-slip at $|z|=R_1$ and $|z|=R_2$. However, at infinite local slip lengths, we obtain the no-shear solution of \citet{Crowdy_2021_theoStud}, since $\beta_1 \rightarrow 1$ and $\beta_2 \rightarrow 0$, representing the ideal limit of an non-viscous interface interaction. As mentioned, the quantity of the local maximum slip length $\tilde{\lambda}$ is to be determined with respect to the system under consideration. The flow field of the pipe or annulus can therefore be calculated by specifying the local maximum slip length imposed by a microstructure, which is dependent on the properties of the enclosed fluid and the groove/structure geometry \citep{ybert2007achieving, Schoenecker_2014_Influenceenclosedfluid}.

\subsection {Flow fields for the pipe and annular flow}
\label{subsec:results-discussion_flow-field}
\begin{figure}
    \centerline{\includegraphics{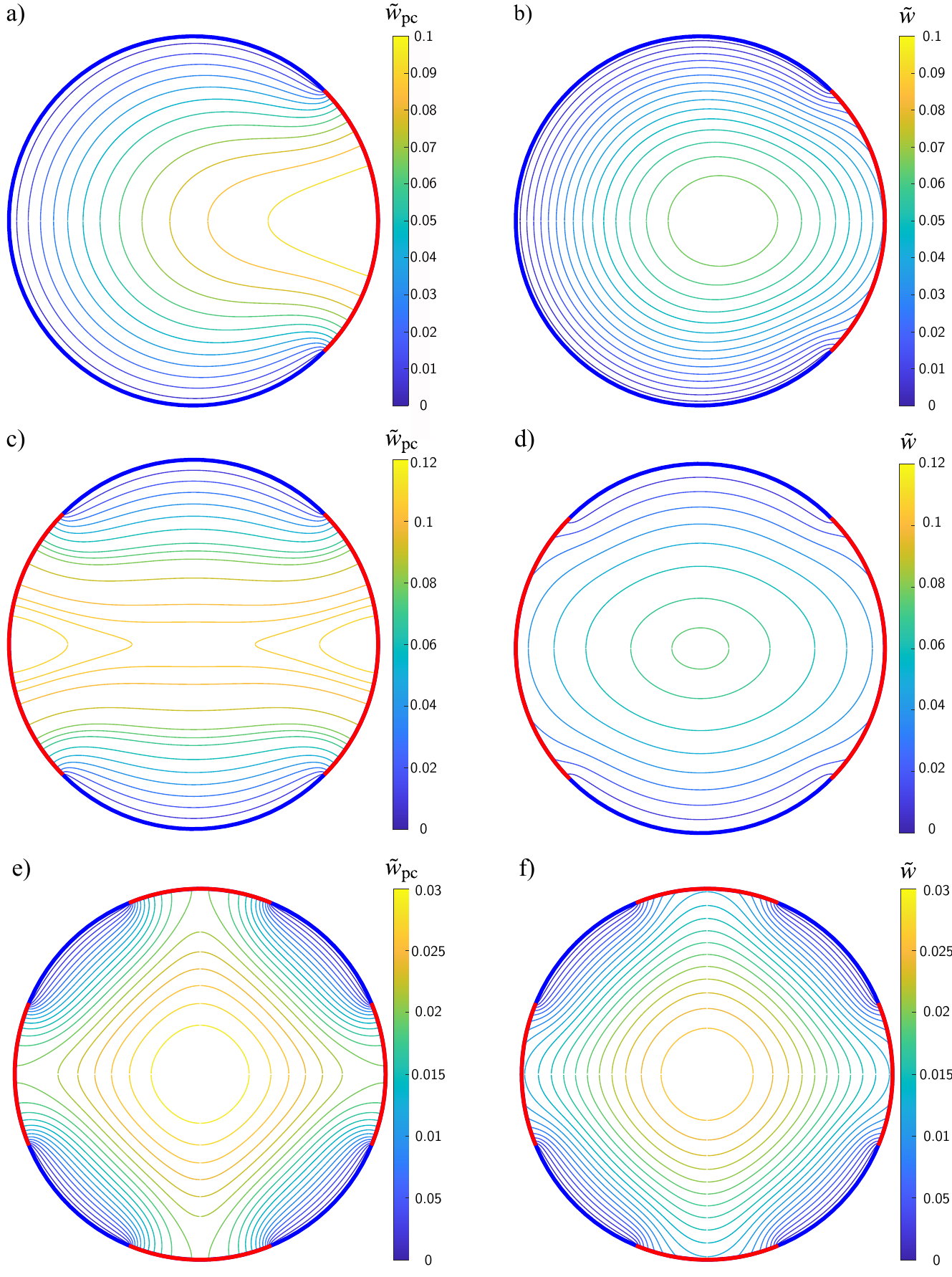}}
    \caption{Variation of velocity contour lines through a pipe along rotationally symmetric slits (red) and no-slip boundaries (blue). Comparing no-shear (normalized eq. \ref{eq:pipeflow_solution_no-shear}) with finite-shear (from eq. \ref{eq:philip-pipe-flow_finite-shear_norm}) solution. a) No-shear slit: $N=1, R=0.5, \theta=\upi/4$. b) Finite shear slit: as a), with $\tilde{\lambda}=0.2$. c) No-shear slits: $N=2, R=0.5, \theta=\upi/2$. d) Finite shear slits: as c), with $\tilde{\lambda}=0.2$. e) No-shear slits: $N=4, R=0.3, \theta=\upi/2$. f) Finite shear slits: as e), with $\tilde{\lambda}=0.2$.}
    \label{fig:contour-lines_pipe-flow_no-shear_shear}
\end{figure}

\begin{figure}
    \centerline{\includegraphics{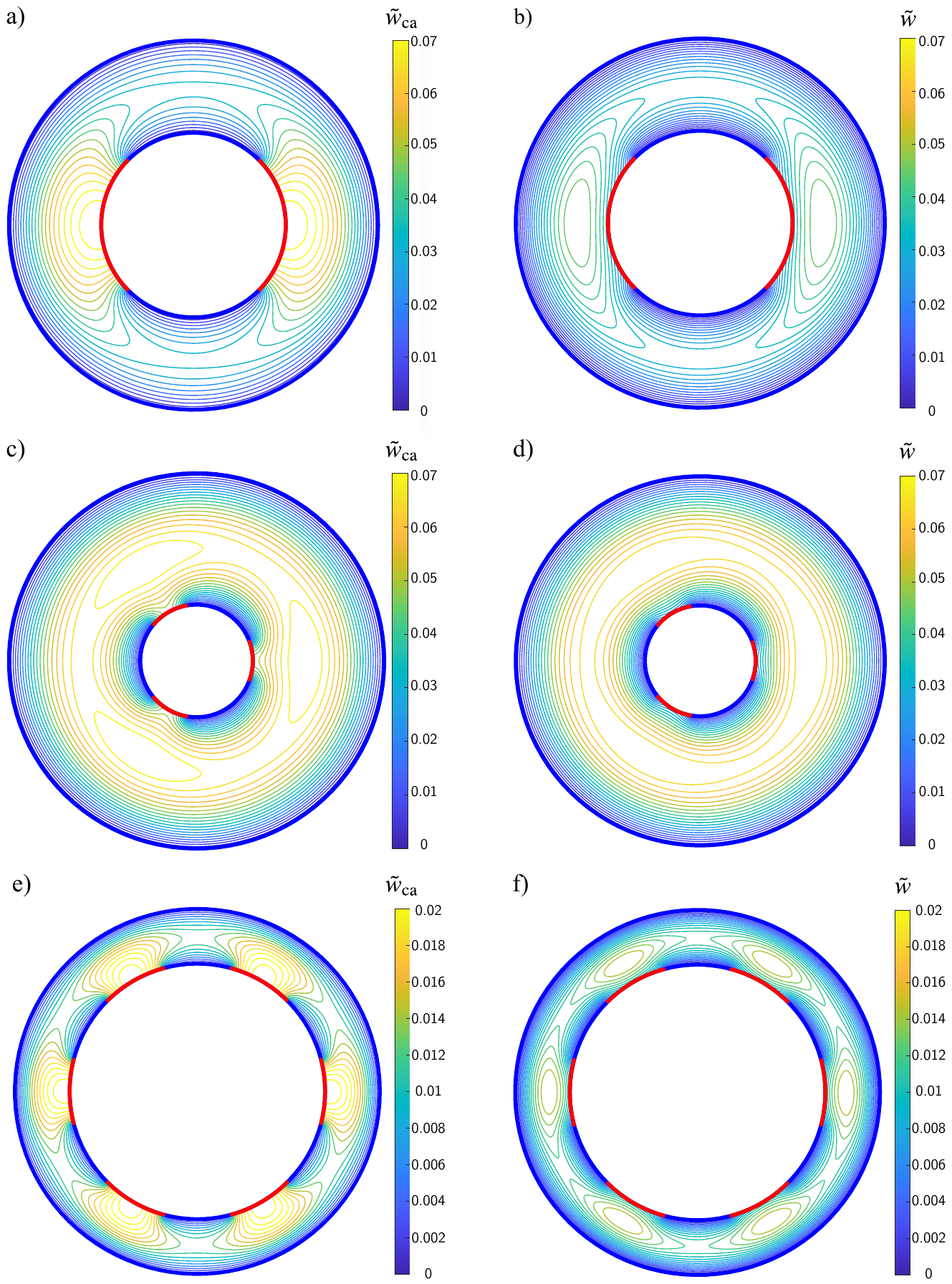}}
    \caption{Variation of velocity contour lines through an annular pipe along rotationally symmetric slits (red) and no-slip boundaries (blue). Comparing no-shear (normalized eq. \ref{eq:annular-pipeflow_solution_no-shear}) with finite-shear (from eq. \ref{eq:annular-pipeflow_solution_finite-slip_norm}) solution. a) No-shear slits: $N=2, R=0.5, \theta=\upi/2$. b) Finite shear slits: as a), with $\tilde{\lambda}=0.2$. c) No-shear slits: $N=3, R=0.3, \theta=\upi/3$. d) Finite shear slits: as c), with $\tilde{\lambda}=0.1$. e) No-shear slits: $N=4, R=0.7, \theta=\upi/2$. f) Finite shear slits: as e), with $\tilde{\lambda}=0.1$.}
    \label{fig:contour-lines_annular-flow_no-shear_shear}
\end{figure}

For the pipe flow solution, examples of velocity contour lines are illustrated in figure \ref{fig:contour-lines_pipe-flow_no-shear_shear}. The left column shows contour plots of the axial velocity, assuming ($N=1,2,4$) no-shear slits along the pipe wall (red), at $|z|=R$. The right column of figure  \ref{fig:contour-lines_pipe-flow_no-shear_shear} corresponds to the same pipe geometries, however, with finite local slip lengths along the grooves. It is easy to see that a finite slip length reduces the overall velocity of the flow compared to the no-shear solution, as expected. For $\tilde{\lambda} \rightarrow \infty$, the right column transitions into the left, consequently corresponding to the ideal limit of non-viscous fluid-fluid interface interaction. 
Figure \ref{fig:contour-lines_annular-flow_no-shear_shear} shows contour plots of normalized pressure driven annular flow solutions, containing $N=2,3,6$ slits (red) at $|z|=R_1$ and no-slip walls (blue) along the remaining boundaries. Whereas the left column assumes no-shear grooves according to Crowdys solution, the right features finite slip lengths. 

\subsection {Connected flow field}
\label{subsec:results-discussion_connec}
\begin{figure}
    \centerline{\includegraphics{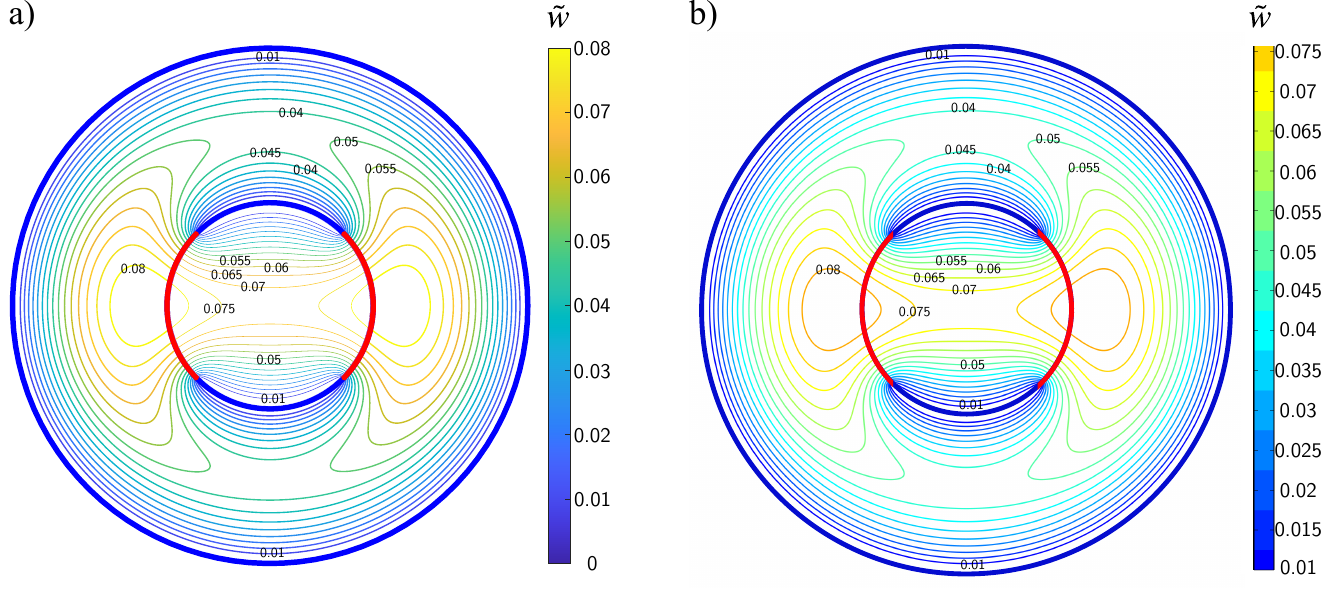}}
    \caption{Illustration of the normalized velocity contour lines with $N=2, R=0.4$ and $\theta=\upi/2$. Assuming $\mu_a = \mu_b$ and $s_a=s_b$.  a) Analytically calculated flow field, from Equation \ref{eq:philip-pipe-flow_finite-slip_regime-a} and \ref{eq:annular-pipeflow_solution_finite-slip_regime-b}. b) Numerically calculated streamlines are shown for comparison. }
    \label{fig:contour-lines_connected-flow_comp_ana_num_s_mu}
\end{figure}

Now we will consider cases where the inner pipe flow is connected to the outer annular flow field and vice versa. Figure \ref{fig:contour-lines_connected-flow_comp_ana_num_s_mu} illustrated such a case for two slits, an inner radius of $R_1 = 0.5$ and $\theta=\upi/2$. In the present example it is assumed that $\mu_a = \mu_b$ and $s_a=s_b$, which would correspond to the case of both pipe flows containing the same fluid and are driven by an equal pressure gradient. The analytically calculated coupled flow field is shown in figure \ref{fig:contour-lines_connected-flow_comp_ana_num_s_mu}(a). 
To verify the derived analytic solutions, numerical calculations have been performed with the commercial finite-element solver COMSOL Multiphysics\textsuperscript{\textregistered}. For that, the two-dimensional Poisson equation has been solved for the inner and the outer flow domain, as illustrated in figure \ref{fig:contour-lines_connected-flow_comp_ana_num_s_mu}(b). Both fluid domains are connected at the red boundary parts. It should be noted that the mathematical connection of both flow fields in the numerical calculation is done along the complete fluid-fluid interface. In contrast, the analytical solution is only coupled at one point, the center of the groove at coordinate $\tilde{\mathfrak{z}}$. The inner and outer no-slip walls (blue) impose $\tilde{w}=0$ and are assumed to be infinitely thin, as is the assumption for the connected analytic flow field solution. The triangular mesh was strongly refined along the interface to adequately resolve the connection conditions, resulting in $374752$ mesh elements. Figure  \ref{fig:contour-lines_connected-flow_comp_ana_num_s_mu}(b) shows the numerically calculated velocity contour plot. The comparison in figure \ref{fig:contour-lines_connected-flow_comp_ana_num_s_mu} clearly shows the excellent agreement of both flow fields. 
\begin{figure}
    \centerline{\includegraphics{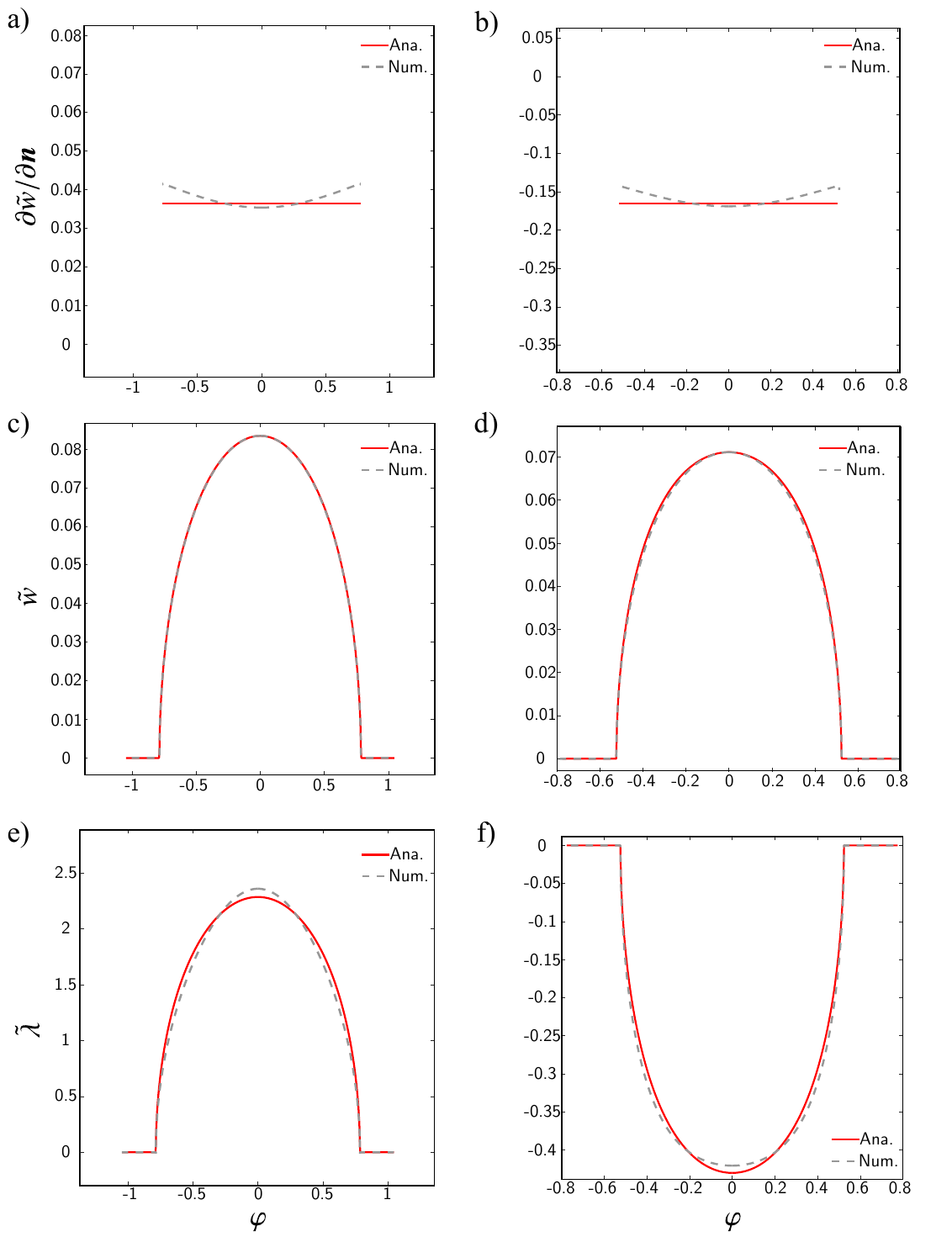}}
    \caption{Comparison of analytically and numerically calculated results along the fluid-fluid interface at $\tilde{z}=R_1 \ \mathrm{exp}(i \varphi)$ with $\varphi \in [-\phi, \phi]$. The analytical solution is based on the combined annular flow field of Equation \ref{eq:annular-pipeflow_solution_finite-slip_regime-b}. Left column: $N=2, R_1=0.4$ and $\theta=\upi/2$. Right column: $N=2, R_1=0.7$ and $\theta=\upi/3$. Assuming $\mu_a=\mu_b$ and $s_a=s_b$ for both cases. a) and b) show the normalized shear stresses, c) and d) the normalized velocities and e) and f) the corresponding local connection slip length distributions.}
    \label{fig:function-plot_comp_vel-shear-slip_ana_num}
\end{figure}

To investigate the agreement more in detail, a comparison of analytically and numerically calculated shear stresses, velocities and the corresponding local connection slip length distributions along the fluid-fluid interface is given in figure \ref{fig:function-plot_comp_vel-shear-slip_ana_num}. All analytical solutions are based on the combined annular flow field of equation \ref{eq:annular-pipeflow_solution_finite-slip_regime-b}. The left column of figure \ref{fig:function-plot_comp_vel-shear-slip_ana_num} examines a pipe-within-pipe flow regime for $N=2, R_1=0.4, \theta=\upi/2$ and the right column for $N=2, R_1=0.7, \theta=\upi/3$, assuming $\mu_a=\mu_b$ and $s_a=s_b$ in both cases. The derived analytical solutions assume constant shear along the interface, as assumed in the ansatz to ensure an explicit solution. The numerical results from figure  \ref{fig:function-plot_comp_vel-shear-slip_ana_num} (a) and (b) show that the shear stress along the interface is not constant. The results differ most at the corners of the grooves. Nevertheless, the comparison for both cases considered shows that the plots match remarkably well. This is especially remarkable since the largest deviations are to be expected along the interface anyway, which further underlines the quality of the assumption made. Analyzing the averaged shear stress along the interfaces further reveals the cumulative error as being rather small, since $\Delta \tilde{\tau} = |\bar{\tilde{\tau}}_{\mathrm{num.}} - \bar{\tilde{\tau}}_{\mathrm{ana.}}| = 0.0377 - 0.0365 = 0.0012$ for the left column and $\Delta \tilde{\tau} = -0.159 - (-0.165) = 0.006$ for the right column in figure \ref{fig:function-plot_comp_vel-shear-slip_ana_num}, with
\begin{equation}
    \bar{\tilde{\tau}} = \frac{1}{2 \phi} \int_{- \phi}^{\phi} \tilde{\tau} \ \mathrm{d}\phi,
\end{equation} 
where $\phi$ still is the half slit angle. The difference can therefore be considered negligible, thus supporting our original assumption of constant shear. 

The agreement between the numerical and analytical results for the normalized velocity along the interface is excellent, as can be seen in figure \ref{fig:function-plot_comp_vel-shear-slip_ana_num} (c) and (d) for both geometries. Examining the plots for the local connection slip length distribution in figure \ref{fig:function-plot_comp_vel-shear-slip_ana_num} (e) and (f), it is readily seen that both curves agree well. The discrepancy of the local connection slip length averaged over the interface
\begin{equation}
    \bar{\tilde{\Lambda}} = \frac{1}{2 \phi} \int_{- \phi}^{\phi} \tilde{\Lambda} \ \mathrm{d}\phi,
\end{equation} 
 is $\Delta \tilde{\Lambda} = |\bar{\tilde{\Lambda}}_{\mathrm{num.}} - \bar{\tilde{\Lambda}}_{\mathrm{ana.}}| = 1.7710 - 1.820 = 0.049$ for the left column and $\Delta \tilde{\Lambda} = -0.350 - (-0.349) = 0.001$ for the right column, respectively. The small observable deviations can be explained by the aforementioned assumption of constant shear stresses at the interface. However, considering the derived flow field solutions being the result of a superposition approach, the agreement is remarkable. 

\begin{figure}
    \centerline{\includegraphics{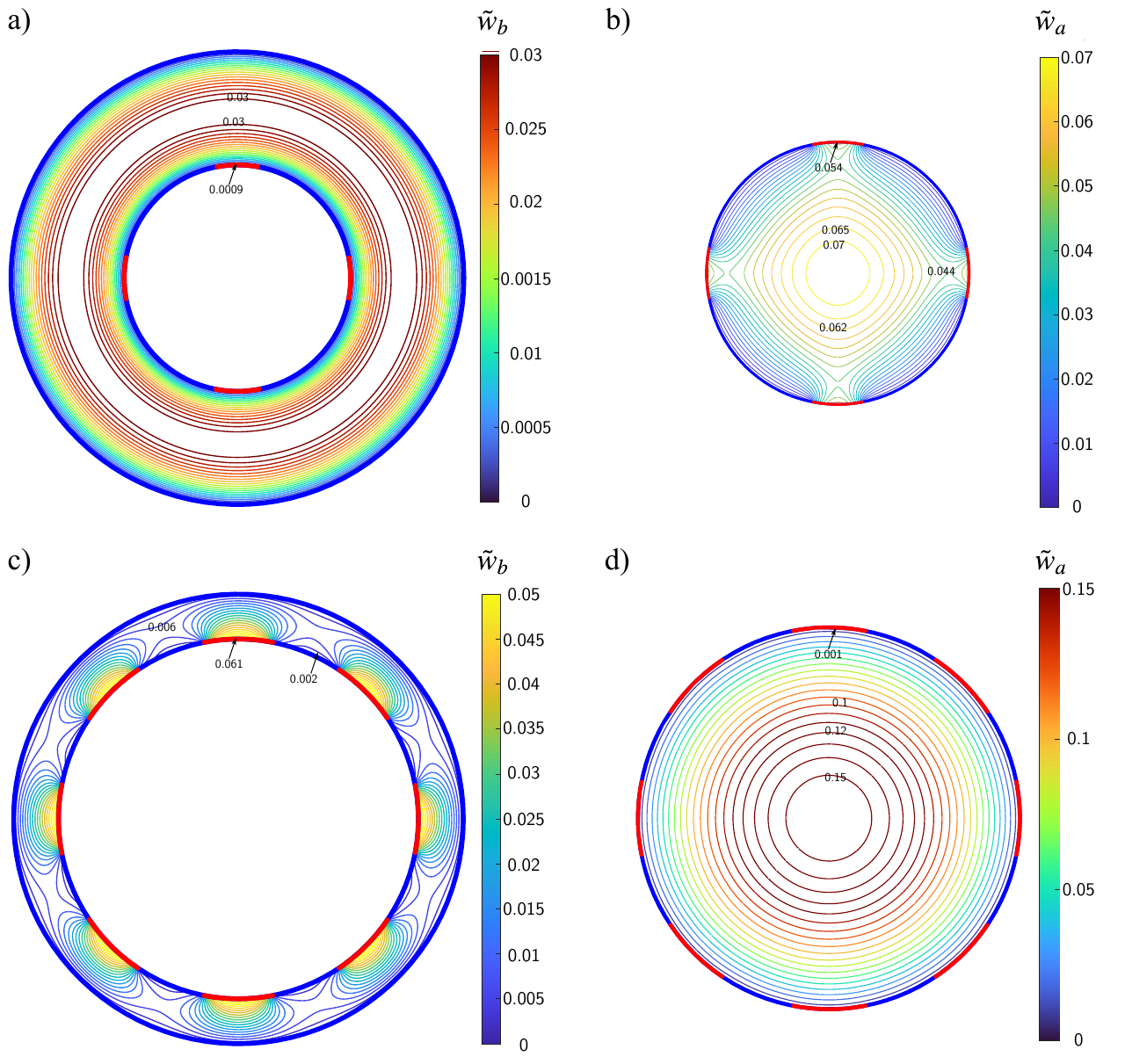}}
    \caption{Illustration of normalized flow field contour lines of a pipe flow (domain a) connected to an annular flow (domain b) with $\tilde{w}_a = w_a \ ( \mu_a/s R_0^2)$, $\tilde{w}_b = w_b \ ( \mu_b/s R_0^2)$ for water ($\mu = \mathrm{10}^{-3}$ Pa s) and air ($\mu =$ 1.8 $\times \ \mathrm{10}^{-5}$ Pa s). a) Air flow through an annular pipe for $N=4, R_1 = 0.5$ and $\theta=\upi/4$. b) Water flow through a pipe connected to a). c) Water flow through an annular pipe for $N=8, R_1 = 0.8$ and $\theta=\upi/2$. d) Air flow through a pipe connected to c). }
    \label{fig:contour-lines_flow-field_water_air}
\end{figure}

\begin{figure}
    \centerline{\includegraphics{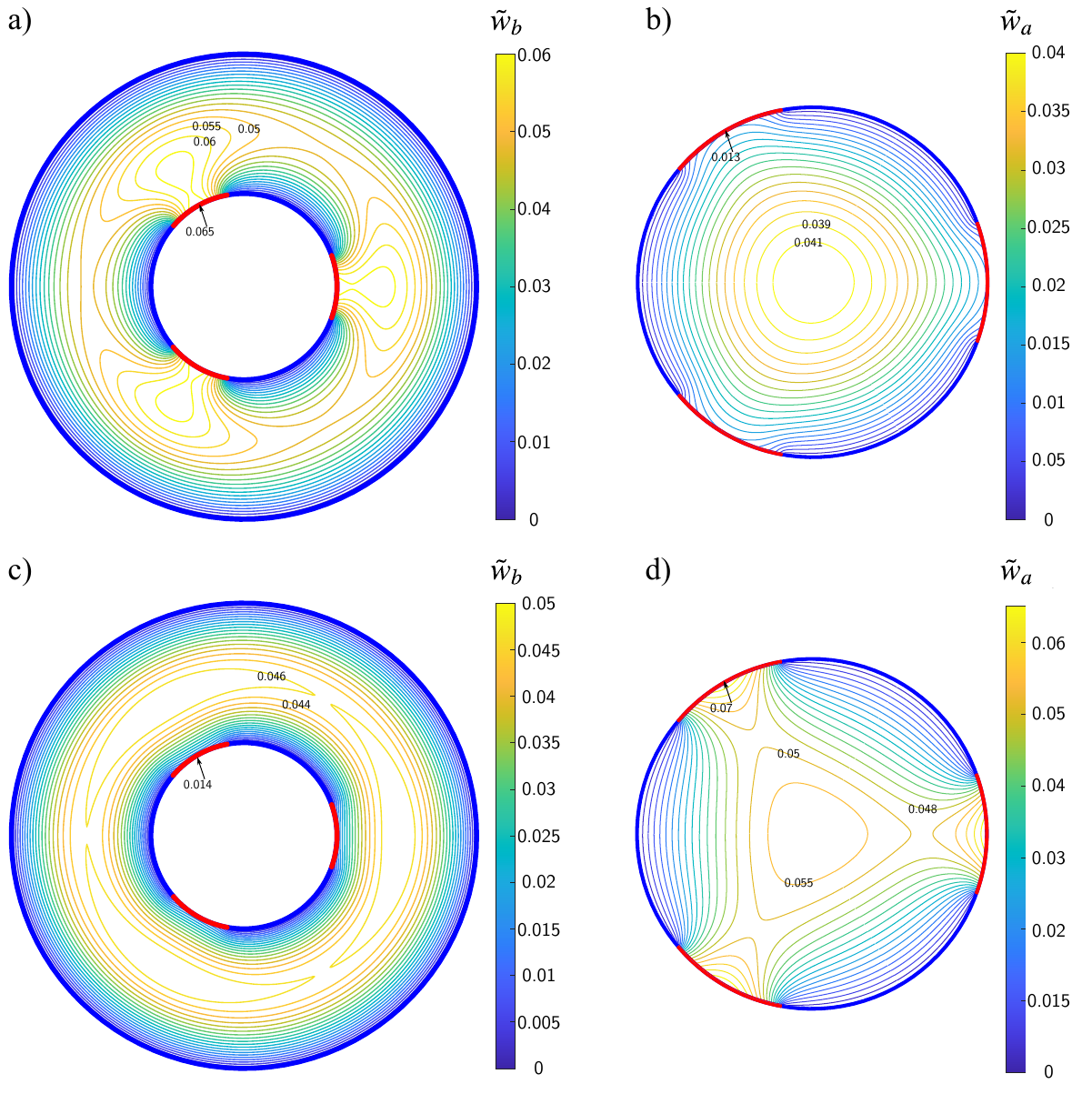}}
    \caption{Illustration of normalized flow field contour lines of a pipe flow (domain a) connected to an annular flow (domain b) with $\tilde{w}_a = w_a \ ( \mu_a/s R_0^2)$, $\tilde{w}_b = w_b \ ( \mu_b/s R_0^2)$ for water ($\mu = \mathrm{10}^{-3}$ Pa s) and oil ($\mu =$5 $\times \ \mathrm{10}^{-3}$ Pa s). Note: For better visibility, the illustrations in the right column are shown enlarged. a) Oil flow through an annular pipe for $N=3, R_1 = 0.4$ and $\theta=\upi/3$. b) Water flow through a pipe connected to a). c) Water flow through an annular pipe for $N=3, R_1 = 0.4$ and $\theta=\upi/2$. b) Oil flow through a pipe connected to a).}
    \label{fig:contour-lines_flow-field_water_oil}
\end{figure}

The previous consideration has been limited to the case $\mu_a = \mu_b$ and $s_a=s_b$. The interaction of air and water is shown in figure \ref{fig:contour-lines_flow-field_water_air}, but still with the same pressure gradient. Contour plots of the normalized axial velocities are illustrated, with both flow fields in a row being connected to each other. Each velocity is normalized with the corresponding viscosity of the associated domain, so $\tilde{w}_a = w_a \ ( \mu_a/s R_0^2)$ and $\tilde{w}_b = w_b \ ( \mu_b/s R_0^2)$. The first row of figure \ref{fig:contour-lines_flow-field_water_air} shows an air flow ($\mu =$ 1.8 $\times \ \mathrm{10}^{-5}$ Pa s) in the annular pipe connected to a pipe water flow ($\mu = \mathrm{10}^{-3}$ Pa s), with $N=4, R_1 = 0.5$ and $\theta=\upi/4$. Although both flows have the same pressure gradient, it can be clearly observed  that the air is comparatively strongly accelerated by the connection of both domains, resulting in an almost rotationally symmetrical velocity field in the annulus. While not immediately apparent, the velocities at the center of each groove are equal, which can be easily verified by dividing by the viscosity. Figures \ref{fig:contour-lines_flow-field_water_air}(c) and (d) show the corresponding reverse case, with $N=8, R_1 = 0.8$ and $\theta=\upi/2$. Again, the air now in the inner tube is disproportionately accelerated compared to the water in the annulus. 
Now, in figure \ref{fig:contour-lines_flow-field_water_oil}, the interaction of water with oil ($\mu = 5 \times \mathrm{10}^{-3}$ Pa s) is illustrated, each with the same pressure gradient. The associated contour plots of the normalized velocities are illustrated, with both domains being connected at the center of the red boundary parts $\mathfrak{z}$. Again, each velocity is normalized with the corresponding viscosity in its domain and both flow fields in a row being connected to each other. It should be noted, that for better visibility, the illustrations \ref{fig:contour-lines_flow-field_water_oil}(b) and (d) in the right column are shown enlarged. The first row considers a water flow in the outer pipe, oil correspondingly in the inner, with $N=3, R_1 = 0.4$ and $\theta=\upi/3$. The bottom row is geometrically unchanged, but the fluids are reversed. 

\subsection {Local slip length for the connected case}
\label{subsec:results-discussion_conn-slip-length}
The local slip length $\tilde{\lambda}$ at the slit centre represents a certain, yet unknown, groove influence on the bulk pipe or annular flow and is to be determined accordingly. As discussed earlier, this depends on the fluid properties of the enclosed fluid as well as the microstructure geometry. Conventionally, confined grooves without their own pressure gradient are considered. This means that the bulk flow is not additionally accelerated by the groove, but merely slowed down less by the enclosed fluid that is passively dragged along compared to a solid no-slip wall. The considered velocity profile of the primary fluid then has a strictly positive normal derivative (normal vector points into the fluid zone by definition), as illustrated in figure \ref{fig:fig_slip_length}. Assuming a velocity in positive coordinate direction, the Navier-slip boundary condition dictates the local slip length to be defined as positive. 

\begin{figure}
    \centerline{\includegraphics{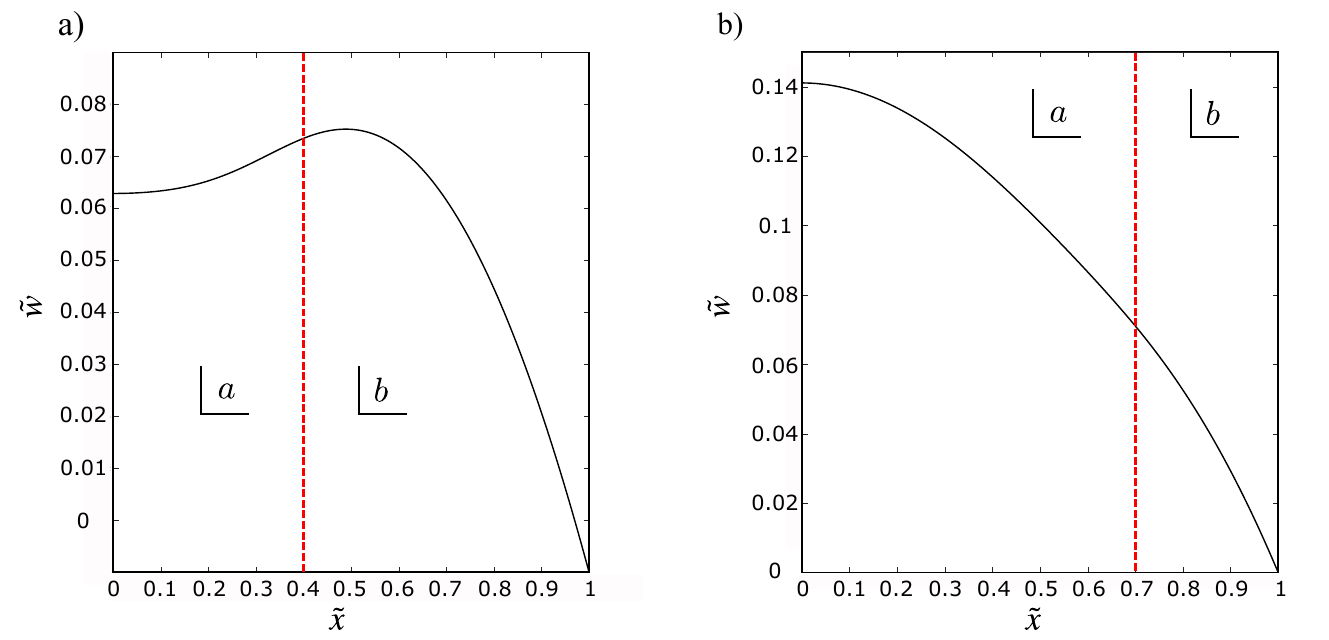}}
    \caption{Change in the normalized axial velocity $\tilde{w}$ as $\tilde{z}$ traverses along the real axis from the centre of the pipe (region $a$) at $\tilde{z}=0+\mathrm{i}0$ to the outer no-slip wall in the annular domain (region $b$) at $\tilde{z}=1+\mathrm{i}0$. The red dashed line marks the transition between the two regions. The intersection of both curves marks the velocity in the centre of the interface $\tilde{z} = \tilde{\mathfrak{z}}$. Assuming $\mu_a = \mu_b$ and $s_a = s_b$. a) $N=2, R_1=0.4$ and $\theta=\upi/2$. b) $N=2, R_1=0.7$ and $\theta=\upi/3$.}
    \label{fig:vel-plots_real-axis}
\end{figure}

The linked pipe-within-pipe scenario is more general than this classical consideration. In the following, by linking the derived pipe flow and annular flow solution of Section 2, mathematical expressions for  $\tilde{\Lambda}_a(w_b)$ and $\tilde{\Lambda}_b(w_a)$ are obtained.  The local connection slip lengths result intrinsically as a function of both individually pressure driven linked flow fields. Both flow regimes can thus experience a sliding effect as well as an added acceleration effect at the fluid-fluid interface due to an additional pressure gradient in the other fluid zone. 
From the perspective of one fluid regime, it is now possible that a higher axial velocity exists beyond the interface. This is obviously not possible with purely shear-driven grooves. Following the previous notation, region $a$ denotes the inner pipe flow and region $b$ the outer annular flow. Figure \ref{fig:vel-plots_real-axis} shows two examples for the velocity in $\tilde{Z}$-direction as one walks along the real axis, starting from the centre of the inner tube to the outer no-slip wall of the annulus. The red dashed line marks the transition between the two regions, i.e. the interface. 
When examining the velocity normal gradient at the interface, there are two cases to distinguish. Considering the condition \ref{eq:pipe-within-pipe_connection-cond_1-2}(b), either 
$$
    \refstepcounter{equation}
    \frac{\partial w_a(\tilde{\mathfrak{z}})}{- \partial \boldsymbol{n}} = - \frac{\partial w_b(\tilde{\mathfrak{z}})}{\partial \boldsymbol{n}} \quad \mathrm{or}\quad \frac{\partial w_a(\tilde{\mathfrak{z}})}{- \partial \boldsymbol{n}} = \frac{\partial w_b(\tilde{\mathfrak{z}})}{\partial \boldsymbol{n}}. 
    \label{eq:case-disting_vel-gradients_connected-flow}
    \eqno{(\theequation \textit{a,b})}
$$
The first case follows from the fact that the velocity gradient of the inner pipe flow is defined in negative radial coordinate direction, contrary to the annular flow part. This means that the normal derivative must always have a different sign in both fluid zones. This becomes evident when looking at e.g. figure \ref{fig:vel-plots_real-axis}(a). The local derivative of the axial velocity profile at the interface of the inner flow is negative as the profile decreases, since the derivative is defined in the negative x-direction ($\partial \tilde{w} / -\partial x$). In contrast, the derivative is positive for the outer flow, since the velocity still increases until reaching a maximum at $\tilde{x} \approx 0.5$. The exact opposite signs arise when considering the case of figure \ref{fig:vel-plots_real-axis}(b).

Whereas equation \ref{eq:case-disting_vel-gradients_connected-flow}(b) only occurs when both velocity gradients are equal at the interface, meaning both must be zero and it therefore must be an inflection point of the axial velocity gradient function. Hence, the shear stress at the interface is also zero.  Locally, this corresponds to the no-shear solution derived by \citet{Philip_1972a} and \citet{Crowdy_2021_theoStud}, respectively. Such a case corresponds to a special case that can be brought about by the appropriate choice of geometry, fluid properties and/or pressure gradients.

It should be noted, that the derived formulas of section \S\ref{sec:coupling} can potentially result in negative local connection slip lengths. This is somewhat unintuitive at first. To clarify, let's consider the Navier-slip condition at the slit center of the case illustrated in figure \ref{fig:vel-plots_real-axis}(a). As established, the velocity profile of the inner pipe flow has a local negative normal derivative at the interface and the outer annular flow a positive one, respectively. At the interface the following must apply
\begin{equation}
    \tilde{\Lambda}_a \underbrace{\frac{\partial w_a(\tilde{\mathfrak{z}}) }{- \partial \boldsymbol{n}}}_{<0} = \tilde{\Lambda}_b \underbrace{\frac{\partial w_b(\tilde{\mathfrak{z}}) }{\partial \boldsymbol{n}}}_{>0}. 
    \label{eq:connection_navier-slip}
\end{equation} 
Since $w_a(\tilde{\mathfrak{z}}) = w_b(\tilde{\mathfrak{z}}) $ must be true at the interface, the local connection slip lengths must correspond accordingly, indicating $\tilde{\Lambda}_a < 0$ and $\tilde{\Lambda}_b > 0$, if $w(\mathfrak{z}) > 0$. In the present example, $\tilde{\Lambda}_a < 0$ thus ensures that the direction of the axial velocity of the internal pipe flow at the interface is in the positive $\tilde{Z}$-direction.  Additionally, negative connection slip lengths can lead to $\alpha(\tilde{\Lambda})>1$. This follows from the simple observation that a flow is driven by the pressure gradient of the linked flow in addition to its own pressure gradient, surpassing the effect of a no-shear boundary. One example would be pipe flow illustrated in figure \ref{fig:contour-lines_flow-field_water_oil}(d), with $\alpha(\tilde{\Lambda}) = 2.34496$. 

The weighting coefficients of the connected annular flow, on the other hand, can additionally be negative. However, this does not result in reverse flow, but may even exceed the case of no-shear slits, which will be shown in the following example. We consider $R_1=0.4, \theta=\upi/3$ and $N=3$. The inner pipe has water ($\mu_a = 0.001 \ \mathrm{Pa} \ \mathrm{s}$) and the annular pipe oil ($\mu_b = 0.005 \ \mathrm{Pa} \ \mathrm{s}$) flowing through it, further assuming $s_1 = s_2$. The weighting coefficients of the annular flow field result  to $\beta_1 = 1.30142$ and $\beta_1 = -0.30142$ accordingly. The associated flow field is shown in figure \ref{fig:contour-lines_flow-field_water_oil}(a). By comparing the velocity in the center of the slits, it can be shown that the present shear example has overall higher velocities than the comparison no-shear solution, since $\tilde{w}(\tilde{\mathfrak{z}}) = 0.065$ and $\tilde{w}_{\mathrm{ca}}(\tilde{\mathfrak{z}}) = 0.049$.

\section {Conclusions}
\label{sec:conclusion}
This article is concerned with modelling pressure-driven pipe flows along longitudinal slits or ridges filled with a second immiscible fluid. In addition to the classical pipe geometry, flow through an annulus containing grooves on its inner wall is also considered. Such flows have numerous applications, including water flowing over superhydrophobic surfaces or along porous media. Surfaces of this type can lead to a significant reduction of flow resistance and are therefore of great importance for the design of energy-efficient technical surfaces. Basis for the corresponding design process is the knowledge of the prevailing flow conditions in the immediate vicinity of the aforementioned microstructures and its consequential influence on the bulk flow. Analytical solutions can decipher these complex dynamics and are therefore of fundamental importance. 

With a superposition technique, analytical approximations for the velocity field of longitudinal flow over rotationally symmetric microstructures in pipes (eq. \ref{eq:philip-pipe-flow_finite-shear_norm}) and annuli (eq. \ref{eq:annular-pipeflow_solution_finite-slip_norm}) were developed. The novelty of these solutions is their dependence on a finite local slip length evaluated at the center of the fluid-fluid interface. 
This contrasts with previous solutions by Philip \citep{Philip_1972a}, Crowdy \citep{Crowdy_2021_theoStud}. Typical solutions to these mixed-valued boundary problems assume perfect interfacial slip, corresponding to an infinite local slip length. This assumption represents an ideal limit of such interface interactions. The extension of these existing approaches performed in this article now allows to factor in the actual influence of individual microstructures. Therefore, viscous interaction along the fluid-fluid interface, flow properties of the enclosed fluid as well as the groove geometry are now taken into account. With the analytical solutions now available, the flow field within a pipe or an annulus can be fully calculated as a function of rotationally symmetric longitudinal slits. The only prerequisite is knowledge of the local slip length at the center of the interface, which remains to be determined on a case-specific basis. The equations are valid for single and multi-phase flow. From these solutions, analytical expressions for the effective slip lengths are determined, which are an important building block in the numerical investigation of superhydrophobic walls.

The derived solutions in section \S\ref{sec:math_description} are important fundamental equations for the calculation of slippery pipes. However, a combination of those results leads to an interesting extension, a pipe-within-pipe geometry (discussed in section \S\ref{sec:coupling}). For that, the pipe and annulus solutions are re-scaled and connected along their interfaces (see figure \ref{fig:connection_pipe_annular_inf-plates_geometry}). This yields two interconnected flow regimes whose physical interfacial communication is modelled using local slip lengths as connection coefficients. Such a pipe-within-pipe promises greater operational control over the slippery effects, as unwanted interface protrusion and potential collapse can be prevented by appropriate external pressure control. The solutions derived in section \S\ref{sec:coupling} are fully determined, i.e. no longer dependent on still unknown local slip lengths. In fact,  these are specified in the connection of both flow regions in dependence of the respective other flow. The resulting flow field equation is then only a function of fluid properties and geometry parameters and therefore easily calculated. 

Overall, a comparison of the analytical solutions with numerical simulations shows excellent agreement. This is particularly remarkable since the derived equations are the result of a comparatively simple superposition approach. This strongly underlines the validity and quality of the mathematical assumption of constant shear stress along all interfaces for such geometries. Ultimately, the presented equations enable further investigation on case-specific optimization of slippage along microstructured pipe walls.

\section*{Acknowledgments}
We kindly acknowledge support by the Deutsche Forschungsgemeinschaft (DFG, German Research Foundation) - Project-ID 467661067. 

\newpage

\bibliographystyle{jfm}
\bibliography{bib-file}

\end{document}